\documentclass[pre,twocolumn,a4paper,showkeys,showpacs]{revtex4}

\usepackage[centertags]{amsmath}
\usepackage{amssymb}
\usepackage{amsfonts}
\usepackage{latexsym}
\usepackage{slashed}
\usepackage{graphicx}
\usepackage{epsfig}
\begin{document}
\title{Bremsstrahlung vs. Thomson scattering in VUV-FEL 
plasma experiments}
\date{\today}
\author{C. Fortmann}
\email{carsten.fortmann@uni-rostock.de}
\homepage{http://www.mpg.uni-rostock.de/vhvi104}
\author{R. Redmer}
\author{H. Reinholz}
\author{G. R\"opke} 
\author{A. Wierling}
\affiliation{Inst. for Physics, Rostock University, 18051 Rostock, Germany}
\author{W. Rozmus} 
\affiliation{Dept. of Physics, University of Alberta, Edmonton, Canada}

\begin{abstract}
\noindent
We determine the spectral photon yield from a hot dense plasma
irradiated by VUV-FEL light in a Thomson scattering experiment. The Thomson signal is
compared to the emission background mainly caused by
bremsstrahlung photons. We determine experimental conditions that allow for
a signal-to-background ratio larger than unity.
By derivation of the Thomson and the bremsstrahlung spectrum from 
linear response theory we present a consistent quantum statistical
approach to both processes. This allows for a systematic treatment of 
medium and quantum effects such as dynamical screening and strong collisions.
Results are presented for the threshold FEL-intensity as a function of density and temperature. 
We show that the account for quantum effects leads to larger thresholds as
compared to previous work.
\end{abstract}
\maketitle
\section{Introduction}
Thomson scattering is a well established technique for experimental investigation of
plasma parameters. Examples can be found in Refs. \cite{sny:pre93_1,snyd:pre93_2,sny:pre94,bent:j.phys.d97,chen:pre02,murp:pre04}. Ob\-ser\-vables like particle density, temperature, composition, and
degree of ionization can be spatially and temporally resolved by analysis of the scattering spectrum
\cite{evan:plph70}. 
Until recently, coherent light sources have been available only for the visible and
near UV part of the electromagnetic spectrum. 
Due to small critical density $n_\mathrm{crit}=\omega^2\epsilon_0m_\mathrm{e}/e^2\approx10^{20}\,\mathrm{cm}^{-3}$ of free charge carriers
for optical probes,
the applicability of Thomson scattering using coherent sources has been
limited to targets of relatively low density.

Glenzer \textit{et al.} \cite{glen:prl03,greg:pre03} have shown and explored the possi\-bi\-lity of x-ray Thomson scattering in 
solid density targets using the Ti He-$\alpha$ line at $4.75\,\mathrm{keV}$ as probe light \cite{greg:phpl04}.
A new alternative emerged with the development of VUV-free electron lasers (VUV-FEL), providing pulses of  
coherent radiation in the far (vacuum-) ultraviolet. At the moment, the VUV-FEL at DESY Hamburg operates at
$32\,\mathrm{nm}$ wavelength \cite{ayva:epjd06}, corresponding to $38\,\mathrm{eV}$ photons. With this coherent light source, dense matter up to 
solid densities of $10^{23}\,\mathrm{cm}^{-3}$  can be penetrated, see Refs.~\cite{hoel:epjd04,redm:ieee05}.
Under these conditions, the Thomson spectrum permits the determination of electron temperature and density directly from
the position and height of collective resonances, i.e. plasmons, showing up in the scattering signal \cite{greg:pre03}.
First experiments will be performed in the  near future 
at the VUV-FEL facility at DESY at $\lambda=32\,\mathrm{nm}$ FEL wavelength,
 while in future stages of the project, wavelengths from 
$13\,\mathrm{nm}$ (VUV-FEL) down to $0.1\,\mathrm{nm}$
(X-FEL) will be available. 

Due to the large number of free charge carriers at the
temperatures and densities considered, 
thermal bremsstrahlung emission, resulting from inelastic 
free-free scattering,
contributes significantly to the 
emission background.
Therefore, experimental conditions such as scattering 
angles, spectral properties of the probe and the detector 
have to be chosen as to obtain a maximum 
signal-to-background ratio.
\textit{Background} is to be understood as bremsstrahlung radiation, whereas \textit{signal} corresponds to the photons having
undergone Thomson scattering.

So far,
classical formulas for 
the bremsstrahlung emission level going
back to Kramers \cite{kram:phil.mag23}  have been used to
determine threshold intensities of the external source
to overcome the background 
due to bremsstrahlung \cite{evan:rep.prog.phys69,bald:rev.sci.inst02}.
The Kramers result, given below in
Eq.~(\ref{eqn:emis_kramers}), is derived from the
assumption of Keplerian
trajectories of the emitting electron in the Coulomb field
of an ion and integration over all initial velocities weighted with
the Maxwellian velocity distribution function.  
Quantum features are only accounted for in a semiclassical way: The velocity integral 
extends over velocities $v$, fulfilling the condition
$mv^2/2\geq \hbar\omega$, i.e. the kinetic energy has to be larger than the
photon energy. Further quantum properties, such as the finite photon momentum as well as the
quantum mechanical nature of the scattering process are not accounted for.
By comparing the Thomson signal strength at the laser wavelength $\lambda=14.7\,\mathrm{nm}$ to the
bremsstrahlung photon yield calculated from Kramers formula, Baldis \textit{et al.} find
threshold intensities of $10^{13}\,\mathrm{W/cm}^2$ for typical values of free electron density $n_\mathrm{e}=10^{22}\,\mathrm{cm}^{-3}$ and 
temperature $k_\mathrm{B}T=100\,\mathrm{eV}$ \cite{bald:rev.sci.inst02}.

In this work, we evaluate threshold
conditions (intensities) using improved expressions for the
bremsstrahlung spectrum.  
As usual, corrections to Kramers formula for bremsstrahlung are described by the so-called
Gaunt factor \cite{gaun:proc.roy.soc30}.
In the simplest
approach it is obtained by taking into account 
collisions between electrons and fixed ions 
in Born approximation.
In dense plasmas, many-particle effects as dynamical screening 
become important. Moreover, strong collisions have to be accounted for.
We show in this paper how the Gaunt factor can be derived from linear response 
theory \cite{zuba2} in a general
way. Within this framework, 
modifications of the emission spectrum beyond Born aproximation 
can be included in a systematic manner \cite{wier:phpl01}
as will be discussed later.

We then apply our formulas to
determine the threshold intensities for a broad range of
experimental parameters (wavelength, spectral properties of
detectors, and different materials), relevant for 
future experiments at DESY.
Furthermore
we compare Thomson and bremsstrahlung photon yield over a finite spectral range.
Thereby, and by taking improved expressions for the bremsstrahlung cross section, 
we show, that even higher thresholds have to be reached in order to obtain a Thomson 
signal above the bremsstrahlung level at least 
near the plasmon resonances. These peaks are much lower than the central peak being essentially an ion feature.

The present work is organized as follows: In the first
section we review the basic physics of Thomson scattering and bremsstrahlung and how they can be expressed in terms of 
the dynamic structure factor and 
the dielectric function, respectively. Since these two quantities are 
related to
each other via the fluctuation-dissipation theorem \cite{rein:pre00}, we are able to describe both processes
on a common and consistent basis.
 
We then compare the emission level due to bremsstrahlung
to the Thomson signal whose strength is proportional to
the flux of incoming photons, i.e. the power density of the
external source. Thereby, we find expressions for
the threshold power density as a function of particle density and temperature. In
the last section we discuss our results for various sets of 
experimental parameters relevant for future experiments at the VUV-FEL.

\section{\label{sec:thomson+brems}Thomson scattering and Bremsstrahlung}
The central quantity of interest is the spectral power density
$\mathrm{d} P/\mathrm{d} V\,\mathrm{d} \lambda\,\mathrm{d} \Omega$, i.e. the rate of 
energy radiated per unit scattering volume $\mathrm{d}V$, wavelength $\mathrm{d}\lambda$, and solid angle $\mathrm{d}\Omega$. 
The total spectral power density is the sum of the corresponding quantity for every radiative
process in the plasma. In this work we focus on Thomson scattering and bremsstrahlung, i.e. 
\begin{equation}
	\frac{\mathrm{d}^3 P_\mathrm{tot}}{\mathrm{d} V\,\mathrm{d}\lambda\,\mathrm{d} \Omega }=
	\frac{\mathrm{d}^3 P_\mathrm{Th}}{\mathrm{d} V\,\mathrm{d}\lambda\,\mathrm{d} \Omega }
	+\frac{\mathrm{d}^3 P_\mathrm{br}}{\mathrm{d} V\,\mathrm{d}\lambda\,\mathrm{d} \Omega }~.
	\label{eqn:P_sum}
\end{equation}
To unambiguously identify the Thomson signal, we require that the Thomson power density
is at least equal to the bremsstrahlung level,
\begin{equation}
	\frac{\mathrm{d}^3P_\mathrm{Th}}{\mathrm{d}V\,\mathrm{d}\lambda\,\mathrm{d}\Omega}\geq 
\frac{\mathrm{d}^3P_\mathrm{br}}{\mathrm{d}V\,\mathrm{d}\lambda\,\mathrm{d}\Omega}~.
	\label{eqn:thresh_cond}
\end{equation}
The Thomson spectrum is given by the intensity of the probe laser $I_\mathrm{L}$, and the Thomson scattering
cross section $\mathrm{d}^2\sigma_\mathrm{Th}/\mathrm{d} \omega\mathrm{d} \Omega$. To account for 
the finite spetral bandwidth of the detector, one has to convolute each power spectrum with a detector function $G(\lambda)$. In practice, this is only
relevant for the Thomson signal since the bremsstrahlung spectrum is 
slowly varying in the relevant frequency region. The Thomson power spectrum reads
\begin{equation}
	\frac{\mathrm{d}^3P_\mathrm{Th}
	(\lambda)}
{\mathrm{d}V\,\mathrm{d}\lambda\,\mathrm{d}\Omega}= 
I_\mathrm{L}\int\mathrm{d}\bar\lambda
G(\lambda-\bar\lambda)
\frac{\mathrm{d}^2\sigma_\mathrm{Th}
(\omega_{\bar\lambda})}
{\mathrm{d}\Omega\,\mathrm{d}\omega}
= I_\mathrm{L}\bar R(\lambda)~.
	\label{eqn:sig_thomson}
\end{equation}

We have introduced the response function $\bar R(\lambda)$, where the bar denotes the convolution with the detector function.
Note that we assume an optically thin plasma, thus radiation transport is neglected.
Also, due to the short pulselength of the VUV-FEL (20-120 fs), we neglect the 
heating of the plasma due to the probe beam.

The bremsstrahlung spectrum does solely depend on the plasma parameters density and
temperature, so that, 
with a suitable formula for the bremsstrahlung spectrum, which we will abbreviate
by the notation $\mathrm{d} ^3P_\mathrm{br}/\mathrm{d} V\,\mathrm{d} \lambda\,\mathrm{d}\Omega\equiv j(\lambda)$ in
the following, Eq.~(\ref{eqn:thresh_cond}) defines a threshold intensity
\begin{equation}
	I_\mathrm{thresh}(\lambda)=\frac{j(\lambda)}{\bar R(\lambda)}~.
	\label{eqn:thresh_int}
\end{equation}
We will now briefly describe how expressions for the Thomson scattering and the bremsstrahlung spectrum
can be obtained from a common starting point, i.e. the dielectric function of the plasma.

\subsection{Thomson scattering}
The cross section for Thomson scattering in a plasma can be given
in terms of the dynamic structure factor (DSF), $S(\boldsymbol{k},\omega)$:
\begin{equation}
	\frac{\mathrm{d}^2
	\sigma_\mathrm{Th}(\boldsymbol{k},\omega)}{\mathrm{d} \Omega\mathrm{d} \omega}=
	\left(\frac{\mathrm{d}\sigma(\Omega)}{\mathrm{d} \Omega}\right)_\mathrm{Th}
	\frac{k_1}{k_0}S(\boldsymbol{k},\omega)~,
	\label{eqn:thoms_skw}
\end{equation} 
see Ref.~\cite{chih:j.phys87,chih:j.phys00} for details. 
The variables $(\boldsymbol{k},\omega)$ are related to the transferred linear momentum and energy, respectively, while $k_0=\omega_0/c$ and
$k_1=|\boldsymbol{k_0}+\boldsymbol{k}|$ are incoming and outgoing linear momenta of the laser field, respectively.
$(\mathrm{d} \sigma(\Omega)/\mathrm{d} \Omega)_\mathrm{Th}$ is the Thomson scattering cross section for the
isolated scattering event. It is given by the Klein-Nishina formula, its derivation can be found in standard textbooks of
quantum electrodynamics, e.g. Ref.~\cite{itzy}. 

In the nonrelativistic limit, the Thomson cross section for unpolarized light is given by
\begin{align}
	\left(\frac{\mathrm{d}\sigma}{\mathrm{d} \Omega}\right)_\mathrm{Th}=\frac{r_\mathrm{e}^2}{2}\left( 1+\cos^2\theta
	\right)~.
	\label{eqn:klein_nishina_kto0}
\end{align}
Here, $r_\mathrm{e}=e^2/4\pi\epsilon_0m_\mathrm{e}c^2$ is the classical electron radius, $\theta$ is the scattering angle
between by $\boldsymbol{k}_1$ and $\boldsymbol{k}_0$.

Equation~(\ref{eqn:thoms_skw}) shows that plasma collective properties can be 
measured in a Thomson scattering experiment. 
However, this requires an accurate theory of the
dynamic structure factor.
The dynamic structure factor $S(\boldsymbol{k},\omega)$ is closely related to the longitudinal dielectric
function by the fluctuation-dissipation theorem \cite{ichi1}
\begin{eqnarray}
\label{eqn:fdt_eps}
S(\boldsymbol{k},\omega) = -\frac{\epsilon_0\hbar k^2}{\pi e^2 n_\mathrm{e}} \,
\frac{\mbox{Im}\,\epsilon_{l}^{-1}(\boldsymbol{k},\omega)}%
{1-\exp\left( -\hbar\omega /k_{\rm B} T \right)} ~.
\end{eqnarray} 
This relation can be utilized by applying an appropriate approximation to the
dielectric function. 
Due to the mass ratio $m_\mathrm{e}/m_\mathrm{i}\ll 1$, Thomson scattering on ions 
can be neglected. Therefore, we will approximate $S(\boldsymbol{k},\omega)$ by 
$S_\mathrm{ee}(\boldsymbol{k},\omega)$, the electronic DSF. 

\subsection{Bremsstrahlung}
Radiative free-free transitions of electrons, known as bremsstrahlung, 
represent the main source of emission in a hot, fully ionized plasma.
However, the bremsstrahlung process requires the presence of a scattering partner that 
carries the recoil momentum. Kinematically, emission by electrons which scatter on ions is dominant.

For a thermally equilibrated plasma, emission,
characterized by the spectral power density $j(\omega)$,
and the absorption coefficient $ \alpha( \omega)
$ as the relative attenuation of 
the intensity of electromagnetic waves propagating in 
the medium per unit length,
are linked by Kirchhoff's law \cite{grie}
$j( \omega)=L( \omega) \alpha( \omega)$,
with the Planck distribution $L( \omega)=\hbar\omega ^3/\left[4\pi^3c^2 (\exp( \hbar\omega/k _{\mathrm{  B}}T)-1)\right]$.  
Note that we consider the long wavelength limit, thus the wavevector $\boldsymbol{k}$ does not show up as an 
argument of the quantities $j(\omega)$ and $\alpha(\omega)$.
The absorption coefficient can be
obtained from the imaginary part of the transverse dielectric function according to
\begin{equation}
  \label{eqn:alpha_eps}
  \alpha(\omega)=\frac{\omega}{c} \frac{{\rm Im}\, \epsilon_t(\omega)}{n(\omega)}~,   
\end{equation}
where the index of refraction $n(\omega)$ is also linked to the transverse dielectric
function by
\begin{equation}
  \label{eq:n_eps}
  n(\omega) = 
  \frac{1}{\sqrt{2}}\left({\rm Re }\, \epsilon_t(\omega) + \left| \epsilon_t(\omega) \right|
  \right)^{1/2}~.
\end{equation}
In the long wavelength limit, the longitudinal dielectric function, which appears in the fluctuation-dissipation 
theorem (\ref{eqn:fdt_eps}), and the transverse dielectric function coincide \cite{maha}. Thus, 
relations (\ref{eqn:fdt_eps}) and (\ref{eqn:alpha_eps}) enable us to treat both
Thomson scattering and bremsstrahlung on a common basis, namely by using an appropriate theory
for the longitudinal dielectric function.

\subsection{\label{subsec:cons_approx}Consistent approximations}
The aim of this work is to compare both radiation processes, Thomson scattering
and bremsstrahlung, calculated in a consistent approximation. 
The comparison has to be carried out between the contributions of either process in
leading order of density.

Bremsstrahlung occurs in 
second order of the coupling constant $\alpha_\mathrm{QED}=1/137$ as can be seen from the transition amplitude $w^\mathrm{br}_\mathrm{fi}$
expressed by Feynman diagrams, 
\begin{equation} w^\mathrm{br}_\mathrm{fi}=\parbox{80pt}{%
	\psfig{width=80pt,figure=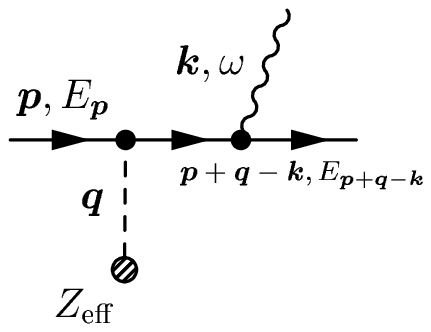}}\qquad+ \qquad\parbox{80pt}{%
	\psfig{width=80pt,figure=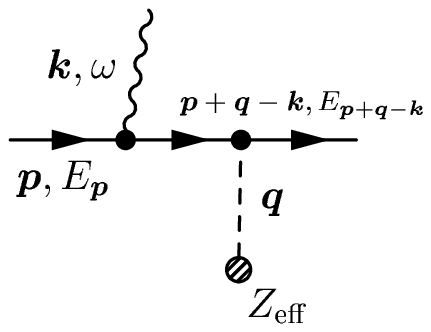}}~,  
	\label{eqn:transampl_feyn} 
\end{equation}
giving the Bethe-Heitler cross section, see Ref.~\cite{itzy}.
The transition amplitude involves a longitudinal field (Coulomb field) 
i.e. a scattering partner, say
an ion of effective charge $Z_\mathrm{eff}$. Thus, bremsstrahlung is naturally of second order in density. A free electron does not emit bremsstrahlung, unless collisions 
take place. The challenge is then to  accuratly describe the scattering process itself.

Born approximation, as given by Eq.~(\ref{eqn:transampl_feyn}) does not describe the correct behaviour of the 
bremsstrahlung spectrum in the case of collisions involving high transfer momenta, so-called strong scatterings. These can be included by ladder-summation 
of all one-photon exchange processes, which leads to the t-matrix. Thereby, the electron-ion interaction is treated accuratly in all orders.
In the Coulomb limit, one finds Sommerfeld's expression for
bremsstrahlung, see Eq.~(\ref{eqn:somm1}) below, and Ref.~\cite{somm1}. On the other hand, due to the long range behaviour of the Coulomb potential, Born approximation 
as well as the Sommerfeld result suffer from infrared divergencies. Here, the account for screening leads to convergent results.
For details, we refer to Ref.~\cite{wier:phpl01}. In our calculations we will use Sommerfeld's formula Eq.~(\ref{eqn:somm1}). Since
we are interested in the bremsstrahlung spectrum at high frequencies, i.e. in the vicinity of the FEL frequency ($\hbar\omega_0=38\,\mathrm{eV}$), no screening effects
are considered here.
Note that the use of the t-matrix does not increase the order in density, instead, it gives the
accurate Gaunt factor in leading order of density. Further many-particle effects like
self-energy and vertex corrections lead to higher order contributions in density and, therefore, are 
not considered here.
For a consistent treatment of the self-energy of the scattering electron, given by multiple scattering on ions and its impact 
on the bremsstrahlung spectrum, see Ref.~\cite{fort:cond.mat.theo05}. 

In the case of Thomson scattering, we adopt the same strategy: We calculate the contribution from lowest order in 
density. Unlike bremsstrahlung, already the first order in density gives a finite
contribution. The amplitude for Thomson scattering in second order of the interaction
\begin{equation} w_\mathrm{fi}^\mathrm{Th}=\parbox{80pt}{%
	\psfig{width=80pt,clip,figure=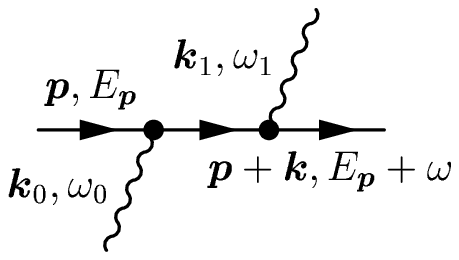}} \qquad+\qquad%
	\parbox{80pt}{%
	\psfig{width=80pt,clip,figure=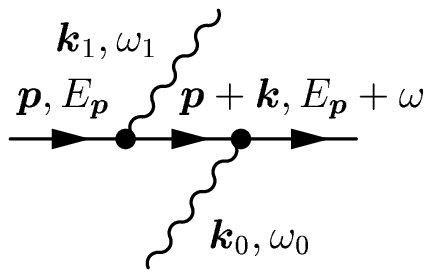}}~, 
	\label{eqn:transampl_thomson_diagr} 
\end{equation}
does not contain any longitudinal field and therefore behaves regularly over the whole spectral range.

Thus,
the DSF,
which describes the Thomson scattering in the medium, 
is taken in random phase approximation (RPA) that accounts for the dynamical screening.
This is all the more important, as the Thomson scattering cross section is evaluated in the
vicinity of the plasma frequency where dynamical screening is the dominant effect \cite{rein:pre00}.
Collisions are neglected, since they occur 
in the next to leading order in density. 

To conclude, a consistent comparison between Thomson scattering and bremsstrahlung in media asks for the
cross section of either process in leading order in density. This consideration leads us to the RPA for the
dynamic structure factor, and a t-matrix ladder summation
for the bremsstrahlung spectrum. 

\section{Linear Response Theory}
Thomson scattering and bremsstrahlung can be 
determined by the longitudinal dielectric function $\epsilon_l(\boldsymbol{k},\omega)$.
We will now briefly outline how the longitudinal dielectric function can be obtained in the framework of 
linear response theory. A discussion of its fundamental aspects
can be found in \cite{zuba2}, for various applications to optical properties of 
plasmas, see \cite{rein:pre00}.
In general, the longitudinal dielectric function $\epsilon_l(\boldsymbol{k},\omega)$ is given in
terms of the dielectric susceptibility
$\chi_{cc'}(\boldsymbol{k},\omega)$, where $c$ and $c'$ label the different species in the plasma and further quantum 
numbers such as spin, 
\begin{equation}
	\label{eqn:eps_chi}
	\epsilon_l(\boldsymbol{k},\omega)
	= \frac{1}{1-\sum_{cc'}
	V_{cc'}(\boldsymbol{k})\chi_{cc'}(\boldsymbol{k},\omega)} ~.
\end{equation}
$V_{cc'}(\boldsymbol{k})$ is the unscreened Coulomb potential between particles of species $c$ and $c'$,
\begin{equation}
	V_{cc'}(\boldsymbol{k})=\frac{q_cq_{c'}}{\epsilon_0k^2}~.
	\label{eqn:coulomb-potential}
\end{equation}

Within linear response theory, the Kubo formula relates the
susceptibility (response 
function) to the current-current correlation function \cite{maha}
\begin{eqnarray}
  \label{eq:Kubo}
  \chi^{}_{cc'}({\boldsymbol k},\omega) & = &  i \beta \Omega_0 
  \frac{k^2}{\omega q_c q_{c'}}\,
  \langle J_{k,c}^z; J_{k,c'}^z \rangle_{\omega+i  \eta}~,
\end{eqnarray}
where $\beta=1/k_\mathrm{B}T$ is the inverse temperature,  
$\Omega_0$ is a normalization volume.
The current operator for particles of species $c$ is defined as
\begin{equation}
	\boldsymbol{J}_{k,c}=\frac{1}{\Omega_0}\sum_{\boldsymbol{p}}\frac{q_c}{m_c}\hbar\boldsymbol{p}a^\dagger_{\boldsymbol{p}-\boldsymbol{k}/2,c}a^{}_{\boldsymbol{p}+\boldsymbol{k}/2,c}~.
\end{equation}

One expresses the current-current correlation function by a force-force correlation function
$\langle\dot J^{z}_{k,c};\dot J^{z}_{k,c'}\rangle_{\omega+i\eta}$, using the time derivative of the
current $\boldsymbol{\dot{J}}_{k,c}=i\big[H, \boldsymbol{J}_{k,c}\big]/\hbar$. The force-force correlation function is more suited for a perturbative treatment, see Ref.~\cite{rein:pre00} and App.~\ref{app:lrt}, where also definitions and useful properties
of correlation functions are given.

\section{Calculation of the bremsstrahlung spectrum}
\begin{figure}[ht]
        \begin{center}
                \psfig{figure=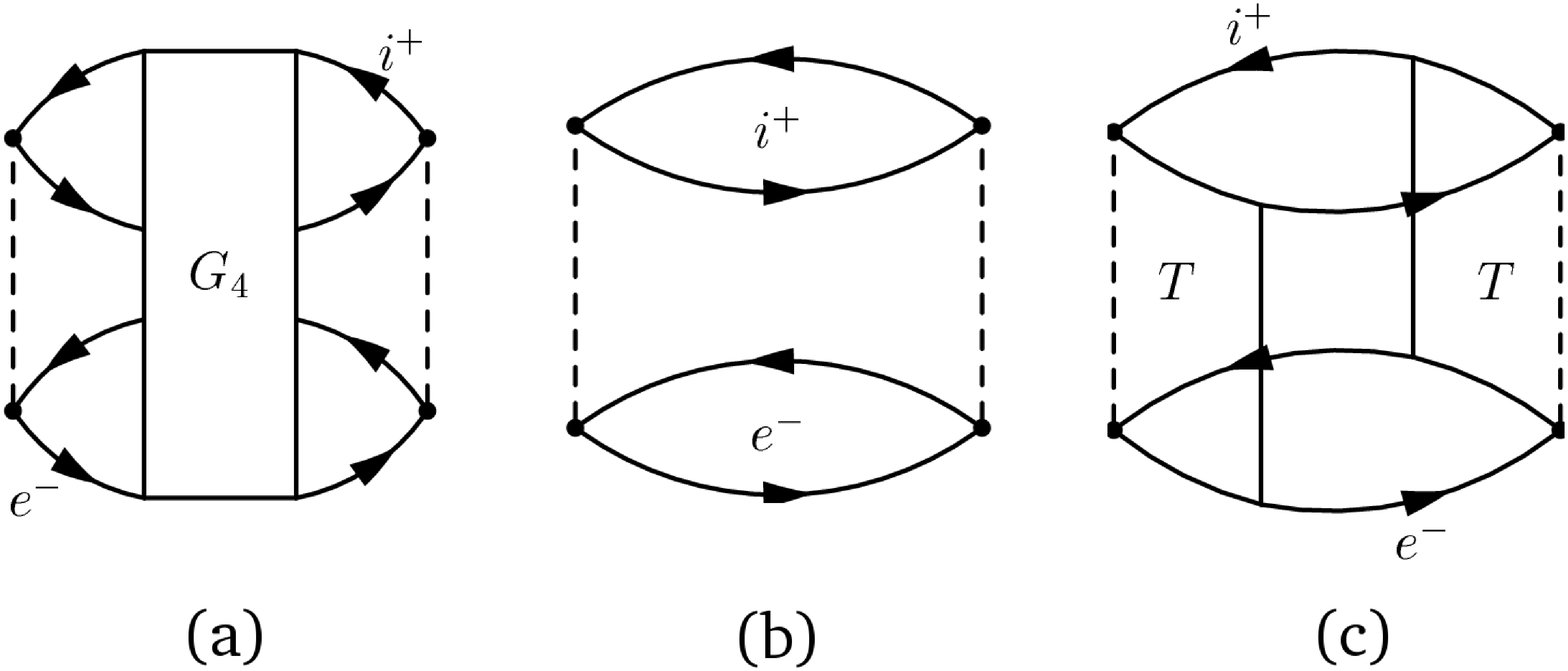,width=.5\textwidth}
        \end{center}
        \caption{Diagrammatic representation of 
	$G_{\dot J \dot J} (\omega_{\mu})$. 
	(a) full account of all
                 medium effects by a four particle Green function,
                 (b) Born approximation, (c) t-matrix
		 approximation.
        	} 
        \label{fig:diag}
\end{figure}

Since the bremsstrahlung spectrum is evaluated at
frequencies far above the plasma frequency,
we can use the high frequency limit of 
Eq.~(\ref{eqn:eps_chi}) in order to derive the
absorption coefficient. For details, we refer to App.~\ref{app:lrt}.
In the high frequency limit, 
the absorption coefficient (\ref{eqn:alpha_eps}) is given by \cite{rein:pre00,wier:phpl01}
\begin{equation}
	\alpha(\omega)=\frac{\beta\Omega_0}{c
	\epsilon_0\omega^2}\mathrm{Re}\,\langle\,\dot
	J_{0,\mathrm{e}}^z,\dot J_{0,\mathrm{e}}^z\rangle_{\omega+i\eta}=\frac{\pi\Omega_0}{c\epsilon_0\omega^3}\mathrm{Im}\,G_{\dot J\dot J}(\omega+i\eta)~.
	\label{eqn:alpha_realjj}
\end{equation}
Here we introduced the force-force 
Green function $G_{\dot J\dot J}(\omega)$, defined in Eq.~(\ref{eqn:def_greenf}),
which can conveniently be calculated using Feynman diagrams, see Fig.~\ref{fig:diag} (a).
$G_4$ denotes a four-particle Green function that contains all
interactions between electrons and ions \cite{wier:phpl01}. In the long wavelength limit considered here, e-e and i-i collisions vanish due
to momentum conservation.
We perform a sequence of approximations by selecting certain
diagrams contributing to $G_4$.

\subsection{\label{born}Born approximation}
As the lowest order contribution with respect to the 
interaction potential, Fig.~\ref{fig:diag} (b) shows the Born approximation.
This approach reproduces the well-known Bethe-Heitler formula in the
nonrelativistic limit:
\begin{align}
	\label{eqn:emis_cl_ln}
	j^\mathrm{B}(\omega)
	&= 
	\frac{16Z_\mathrm{eff}^2n_\mathrm{i}n_\mathrm{e}}{3c^3}
	\left( \frac{e^2}{4\pi\epsilon_0} \right)^3\left( \frac{\beta}{2\pi m_\mathrm{e}} \right)^{3/2}\times\nonumber\\
	&\qquad\qquad\times\int\limits_{\hbar\omega}^\infty 
	\mathrm{d} E^\mathrm{e}_p\,
	\mathrm{e}^{-\beta E^\mathrm{e}_p}\,
	\ln\left|\frac{\sqrt{E^\mathrm{e}_p}+
	\sqrt{E^\mathrm{e}_p-\hbar \omega}}
	{\sqrt{E^\mathrm{e}_p}-
	\sqrt{E^\mathrm{e}_p-\hbar \omega}}\right|
	\\
	&=\frac{8Z_\mathrm{eff}^2n_\mathrm{i}n_\mathrm{e}}{3m_\mathrm{e}c^3\pi}\left( \frac{e^2}{4\pi\epsilon_0} \right)^3
	\sqrt{\frac{\beta}{2\pi m_\mathrm{e}}}
	\mathrm{e}^{-\beta\hbar\omega/2}K_0(\beta\hbar\omega/2)~.
	\label{eqn:emis_cl_K0}
\end{align}
$E^c_{\boldsymbol{p}}$ is the free particle
energy, for nonrelativistic particles
$E^c_{\boldsymbol{p}}=\hbar^2p^2/2m_c$ holds.
Details of the calculation are discussed in
\cite{fort:cond.mat.theo05}.
In Eq.~(\ref{eqn:emis_cl_ln}) the effective ion charge $Z_\mathrm{eff}$ is used to account for the screening of the charge $Ze$ of the nucleus due to 
inner shell electrons. In Ref.~\cite{bald:rev.sci.inst02}, 
$Z_\mathrm{eff}$ is calculated as a function of temperature and density in the framework of Thomas-Fermi theory, see Ref.~\cite{more:ir_llnl} for details. Alternatively, $Z_\mathrm{eff}$ can be obtained in the so-called chemical picture by solving Saha's equation, see Ref.~\cite{kuhl:cpp05}. Especially for high temperatures, the chemical picture gives more
reliable results, as the ionization equilibria between different ionization stages
are satisfied. We will therefore use the chemical picture in this work and
calculate the mean ionization level with COMPTRA04 \cite{kuhl:cpp05}. The Thomas-Fermi
model will only be applied for the purpose of comparison to results obtained by
Baldis \textit{et al.} in Ref.~\cite{bald:rev.sci.inst02}.

The logarithm in Eq.~(\ref{eqn:emis_cl_ln}) is the nonrelativistic limit of 
the Bethe-Heitler cross section \cite{jack,heit}.
The drawback of this result is 
the divergence of the bremsstrahlung spectrum in the limit $\omega\to 0$. However,
physically this is of no importance, since for low frequencies the index of refraction is different from unity and
modifies the spectrum significantly. This is known as dielectric suppression \cite{ter:dokl53}. 

It is common use to write formulas for the emission and
absorption due to (inverse) bremsstrahlung in terms of 
Kramers classical result \cite{kram:phil.mag23}
\begin{equation}
	j^\mathrm{K}(\omega)= \frac{8Z_\mathrm{eff}^2n_\mathrm{e}n_\mathrm{i}}{3m_\mathrm{e}c^3}\left(\frac{e^2}{4\pi\epsilon_0}\right)^3
	\left(\frac{\beta}{6\pi m_\mathrm{e}}\right)^{1/2}
	\exp(-\beta\hbar\omega)~,
	\label{eqn:emis_kramers}
\end{equation}
multiplied with
a correction factor $\bar{g}_\mathrm{ff}(\omega)$, 
called Gaunt-factor \cite{gaun:proc.roy.soc30}, which takes into
account medium effects as well as quantum corrections, 
i.e.
\begin{equation}
	j(\omega)=j^\mathrm{K}(\omega)\cdot
\bar{g}_\mathrm{ff}(\omega)~.
\label{eqn:def_gaunt}
\end{equation}
With Eq.~(\ref{eqn:emis_cl_K0}), the Gaunt factor in Born approximation reads
\begin{equation}
	\bar{g}^\mathrm{B}_\mathrm{ff}(\omega)=
	\frac{\sqrt{3}}{\pi}\exp(\beta\hbar\omega/2)
	K_0(\beta\hbar\omega/2)~.
	\label{eqn:gaunt_K0}
\end{equation}
\subsection{\label{subsec:t-matrix}Strong collisions}
The Born approximation assumes free particles as the in- and out states in
the scattering amplitude. Taking the true scattering states, i.e Coulomb wavefunctions, leads to the so-called Sommerfeld formula \cite{somm1},
\begin{widetext}
\begin{eqnarray}
	\label{eqn:somm1}
	\bar{g}_\mathrm{ff}^\mathrm{S}(\omega,T)&=& \frac{1}{k_\mathrm{B}T}
	\int\limits_0^\infty\mathrm{d}
	E_i\,\mathrm{e}^{-E_i/k_\mathrm{B}T}
	g^\mathrm{S}_\mathrm{ff}(\omega,E_i)~,\\ \nonumber
	\mbox{with}\\
	g^\mathrm{S}_\mathrm{ff}(\omega,E_i)&=&
	\frac{4\sqrt{3}}{\pi}
	\left[ \frac{\left( \eta_i^2+\eta_f^2+2\eta_i^2\eta_f^2
	\right)}{2\eta_i\eta_f}I_0-\left( 1+\eta_i^2
	\right)^{1/2}\left( 1+\eta_f^2 \right)^{1/2}I_1
	\right]I_0~,\label{eqn:somm2}\\\nonumber\\ \nonumber
	\mbox{and}&&\\ \nonumber
	I_l&=& \frac{1}{4}\left[%
	\frac{4\sqrt{E_iE_f}}{\left(%
	\sqrt{E_i}-\sqrt{E_f} \right)^2}
	\right]^{l+1}\left|
	\frac{\sqrt{E_i}-\sqrt{E_f}}{\sqrt{E_i}+\sqrt{E_f}}\right|^{i\left(
	\eta_i
	+\eta_f\right)}\mathrm{e}^
	{\pi\left|\eta_i-\eta_f\right|/2}
	\frac{\left|\Gamma\left(%
	l+1+i\eta_i \right)\Gamma\left( l+1+i\eta_f
	\right)\right|}{\Gamma\left( 2l+2
	\right)}\nonumber\times\\
	&&\times{}_2F_1\left( l+1-i\eta_f,l+1-i\eta_i;
	2l+2;-\frac{4\sqrt{E_iE_f}}{\left(%
	\sqrt{E_i}-\sqrt{E_f} \right)^2}\right)~,\qquad E_f=E_i-\hbar\omega~.
	\label{eqn:somm3}
\end{eqnarray}
\end{widetext}
$\eta_{i/f}^2=Z_\mathrm{eff}^2 \mathrm{Ry}/E_{i/f}$ is the Sommerfeld parameter with the Rydberg energy
$\mathrm{Ry}=m_\mathrm{e}e^4/2(4\pi\epsilon_0\hbar)^2\simeq 13.6\,\mathrm{eV}$,
${}_2F_1(a,b;c;d)$ is the hypergeometric function
\cite{abra} and $\Gamma(x)$ is the Gamma function. 
As shown in Ref.~\cite{wier:phpl01}, Sommerfeld's expression is also obtained by a t-matrix ladder summation with a statically screened Debye potential 
(Fig.~\ref{fig:diag} (c)) if the limit of vanishing inverse screening length (Coulomb limit) is taken.
It gives the correct Gaunt factor in the low density limit, which is considered here. 

\begin{figure}[ht]
	\begin{center}
		\psfig{figure=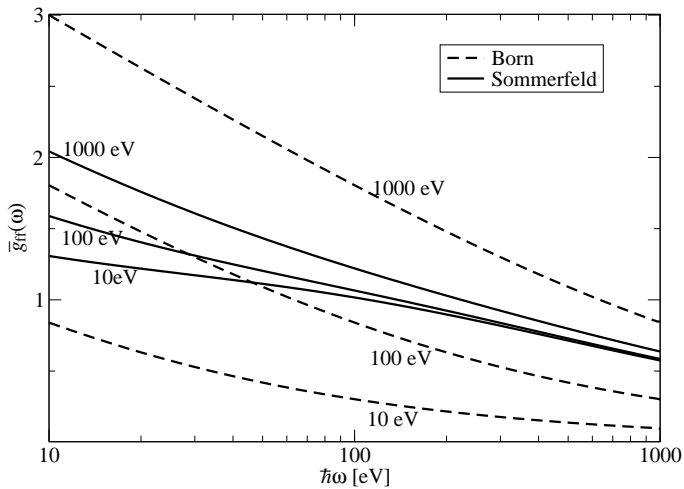,
		width=.5\textwidth,clip,angle=-0}
	\end{center}
	\caption{Gaunt factor $\bar{g}_\mathrm{ff}(\omega
	)$ in Born approximation Eq.~(\ref{eqn:gaunt_K0})
	and using Sommerfeld's formula
	Eq.~(\ref{eqn:somm1}), respectively. 
	Results are presented as a function of 
	the photon energy $\hbar\omega$ for various
	temperatures.}
	\label{fig:gaunt}
\end{figure}

Fig.~\ref{fig:gaunt} shows the dependence of the Gaunt
factor on the photon energy for electron temperatures
$k_\mathrm{B}T=10\,\mathrm{eV},\,100\,\mathrm{eV}\mbox{\ and\ }1000\,\mathrm{eV}$. Over the large
energy interval shown, the Gaunt factor in either 
calculation for a fixed
temperature does not vary much and is of order unity, 
thus the widely used approximation to set 
$\bar{g}_\mathrm{ff}(\omega )\approx 1$ is justified for
estimates of the emission level as done in Refs.~\cite{evan:rep.prog.phys69,bald:rev.sci.inst02}. 
More specific,  $\bar{g}_\mathrm{ff}(\omega)=1$ is a good
approximation for photon energies comparable to
the temperature $\hbar\omega/k_\mathrm{B}T\simeq 1$ 
and for temperatures in the vici\-nity of the
ionization energy ($k_\mathrm{B}T\simeq Z^2\mathrm{Ry}$ for 
hydrogenic systems). 
If one of these conditions is not met, one should use
the full Sommerfeld expression (\ref{eqn:somm1}) or appropriate
approximations, see the detailed discussion in Ref.~\cite{karz:apjs61}.
For the VUV-FEL experiments, both requirements
are satisfied only roughly: The laser provides photon energies of 
$\hbar\omega\simeq 40 \,\mathrm{eV}$ and the optical
pump laser will excite the plasma to 
temperatures of 
$k_\mathrm{B}T\simeq 10\ldots 50\,\mathrm{eV}$, which is
of almost the same order of magnitude as the first ionization
energies of aluminum ($6\,\mathrm{eV},19\,\mathrm{eV},28\,\mathrm{eV}$), or hydrogen ($13.6\,\mathrm{eV}$) 
\cite{allen73}, materials presumably
 used as target in the experiments.
Thus we take into account the Gaunt factor in
t-matrix approximation (Eq.~(\ref{eqn:somm1})) in our calculations.

In Ref.~\cite{wier:phpl01}, a consistent treatment of the impact of dynamical screening and strong collisions on bremsstrahlung based on the Gould-DeWitt scheme \cite{gdw:pr67} is given.
It is shown that the high frequency behaviour of the
Gaunt factor is dominated by the t-matrix contribution, whereas
dynamical screening can be neglected as long as the considered frequencies are large compared to the plasma frequency.


\section{Dynamic structure factor}
\label{sec:skw}
The total dynamic structure factor is defined as
\cite{kubo:jpsj85}
\begin{eqnarray}
\label{Eq:DefStructFac}
S_{cc}(\boldsymbol{k},\omega) = \frac{1}{2 \pi N}\int\limits_{-\infty}^{+\infty} \mathrm{d}t\,
{\rm e}_{}^{i\omega t} \left\langle 
\rho_c(\boldsymbol{k},t)\rho_c(-\boldsymbol{k},0)  \right\rangle ~;
\end{eqnarray}
$\left\langle \dots \right\rangle$ denotes the ensemble equilibrium
average. 
Here, our discussion is focused  on the
electron structure factor $S_\mathrm{ee}(\boldsymbol{k},\omega)$. 
The fluctuation-dissipation theorem (\ref{eqn:fdt_eps}) in connection with Eq.~(\ref{eqn:eps_chi}) allows to 
express the electronic part of the dynamical structure factor via the electronic dielectric susceptibility 
$\chi_\mathrm{ee}(\boldsymbol{k},\omega)$, 
\begin{equation}
	S_\mathrm{ee}(\boldsymbol{k},\omega)=	\frac{\hbar}{\pi n_\mathrm{e}} \,
	\frac{\mbox{Im}\,\chi_\mathrm{ee}(\boldsymbol{k},\omega)}%
	{1-\exp\left( -\hbar\omega /k_{\rm B} T_\mathrm{e} \right)} ~.
	\label{eqn:fdt_chi}
\end{equation}
We will consider the classical limit, where the denominator of Eq.~(\ref{eqn:fdt_chi}) can be approximated by $\hbar\omega/k_\mathrm{B}T_\mathrm{e}$.
In RPA, the following equation for the susceptibility holds \cite{selc:pre01}:
\begin{equation}
	\chi^\mathrm{RPA}_{cc'}(\boldsymbol{k},\omega)=\chi_c^0(\boldsymbol{k},\omega)\delta_{cc'}+\chi_c^0(\boldsymbol{k},\omega)V_{cc'}^\mathrm{sc}(\boldsymbol{k},\omega)\chi_{c'}^0(\boldsymbol{k},\omega)~.
	\label{eqn:chi_rpa_vsc_text}
\end{equation}
$V^\mathrm{sc}(\boldsymbol{k},\omega)$ is the screened
interaction potential which acts between particles of
species $c$ and $c'$. It satisfies the equation
\begin{equation}
	V^\mathrm{sc}_{cc'}(\boldsymbol{k},\omega)=V_{cc'}(\boldsymbol{k})+\sum_d V_{cd}(\boldsymbol{k})\chi_d^0(\boldsymbol{k},\omega)V^\mathrm{sc}_{dc'}(\boldsymbol{k},\omega)~.
	\label{eqn:dyson_Vsc_text}
\end{equation}
The \textit{free} susceptibility
$\chi_c^0(\boldsymbol{k},\omega)$ is obtained from
Eq.~(\ref{eq:Kubo}), taking the current-current correlation
function only in zeroth order with respect to the
interaction, i.e.
\begin{multline}
	\chi_c^0(\boldsymbol{k},\omega)= i \beta \Omega_0 
	\frac{k^2}{\omega q_c q_{c'}}\,
  \langle J_{k,c}^{z}; J_{k,c'}^{z} \rangle_{\omega+i  \eta}^0\\
  = \delta_{cc'}\sum_p\frac{f^c_{\boldsymbol{p}+\boldsymbol{k}/2}-
	f^c_{\boldsymbol{p}-\boldsymbol{k}/2}}
	{E^c_{\boldsymbol{p}+\boldsymbol{k}/2}-
	E^c_{\boldsymbol{p}-\boldsymbol{k}/2}-
	\hbar(\omega+i\eta)}~.
	\label{eqn:chi_0}
\end{multline}
$f^c_{\boldsymbol{p}}$ is the momentum distribution function
of species $c$.  
Again, a small but finite imaginary
frequency $i\eta,\quad \eta>0$ has to be introduced, in order to fix the sign of the imaginary part and
to obtain convergent results.
In the case of a classical two-component plasma with different temperatures $T_c$, we take
the Maxwell distribution.
In the limit $\hbar\to 0$
the resulting susceptibility reads
\begin{equation}
	\chi^\mathrm{0,cl}_c(k,\omega)=	-\Omega_0n_c
	W(x_c)/k_\mathrm{B}T_c~.
	\label{eqn:chi_0_W}
\end{equation}
with the plasma dispersion function
\begin{equation}
	W(x_c)= 1-2x_c\,\mathrm{e}^{-x_c^2}\int\limits_0^{x_c}\,
	\mathrm{e}^{t^2}\mathrm{d}
	t\,-i\sqrt{\pi}x_c\,\mathrm{e}^{-x_c^2}~,
	\label{eqn:wx}
\end{equation}
and the dimensionless variable $x_c=\sqrt{\omega^2 m_c/2k^2k_\mathrm{B}T_c}$.
Eqs.~(\ref{eqn:chi_rpa_vsc_text}) and (\ref{eqn:dyson_Vsc_text}) can be solved algebraically
for the RPA-susceptibilities
$\chi_{cc'}^\mathrm{RPA}(\boldsymbol{k},\omega)$ as shown in App.~\ref{app:chi_rpa}, see also Ref.~\cite{selc:pre01}.  
Evaluation of Eq.~(\ref{eqn:fdt_chi}) with $\chi^\mathrm{RPA}_\mathrm{ee}(\boldsymbol{k},\omega)$, see Eqs.~(\ref{eqn:chi_0}), and (\ref{eqn:im_chi_rpa}), yields the electronic DSF in RPA, i.e.
\begin{multline}
	S_\mathrm{ee}(\boldsymbol{k},\omega )=\\ 	
	\left|\frac{1+Z_\mathrm{eff}\alpha^2(T_\mathrm{e}/T_\mathrm{i})
	W(x_\mathrm{i})}{1+\alpha^2W(x_\mathrm{e})+\alpha^2Z_\mathrm{eff}(T_\mathrm{e}/T_\mathrm{i})
	W(x_\mathrm{i})}\right|^2
	\frac{x_\mathrm{e}\exp\left( -x_\mathrm{e}^2\right)}
	{\omega\sqrt{\pi}}+\\
	+Z_\mathrm{eff}\!\!\left|\!\frac{-\alpha^2W(x_\mathrm{e})}
	{1+\alpha^2W(x_\mathrm{e})+\alpha^2Z_\mathrm{eff}(T_\mathrm{e}/T_\mathrm{i})
	W(x_\mathrm{i})}\!\right|^2\!\!\frac{x_\mathrm{i}\exp\left(-x_\mathrm{i}^2 \right)}{\omega\sqrt{\pi}}.\\
	\label{eqn:skw_evan}	
\end{multline}
Here, the scattering parameter 
\begin{equation}
	\alpha=\kappa_\mathrm{e}/k=\sqrt{n_\mathrm{e}e^2/m_\mathrm{e}k_\mathrm{B}T_\mathrm{e}}/k
	\label{eqn:alpha_def}
\end{equation}
has been introduced. It separates the regime of collective Thomson scattering ($\alpha\gg1$), where collective resonances (plasmons) appear in the spectrum, and the regime of noncollective Thomson scattering ($\alpha\ll1$), where the spectrum reflects the
single-particle distribution function of electrons without further structures.
Expression (\ref{eqn:skw_evan})
was also used by Baldis \textit{et al.} in \cite{bald:rev.sci.inst02}.

Eq.~(\ref{eqn:skw_evan}) for the DSF contains two terms: 
The first term is basically the DSF of a free electron gas. However, it also contains the ionic
dispersion function, which accounts for the screening
effect of ions. 
The second term gives the scattering signal from electrons
which are close to ions and are therefore determined by the
dynamics of these heavy particles. This term is dominant at
small $\omega$ where $W(x_\mathrm{e})$ can be approximated by its
static limit, i.e. $W(x_\mathrm{e}\ll1)\to 1$. For larger values of
$\omega$, the first term dominates the spectrum, because
the
ion part is damped out more rapidly than the electron part.
In particular, plasmon resonances, which appear in the 
spectrum due
to a vanishing real part of the denominator, only survive
in the first term, while they are damped out in the second.
In so-called \textit{Salpeter
approximation} \cite{salp:pr60} the DSF (\ref{eqn:skw_evan}) is separated into two terms 
depending only on electronic and on ionic variables
respectively, electron-ion correlations are neglected to a large extent. Therefore, the full expression 
(\ref{eqn:skw_evan}) should be used, as done throughout this work.

Chihara \cite{chih:j.phys00} gives the DSF in terms of the local field correction factor, thereby including electron-ion collisions.
In appropriate limits, his expression coincides with Eq.~(\ref{eqn:skw_evan}), see Ref.~\cite{chih:j.phys87}. On the other hand,
due to the equivalence of the local field correction and the collision frequency in the high frequency limit, 
Chihara's expression leads to the same formula for the absorption coefficient $\alpha(\omega)$ as given in Eq.~(\ref{eqn:alpha_eps}) \cite{chih:pcFeb06}. 
Thus, Chihara presents
an alternative approach to the question of photoabsorption and emission and Thomson scattering in plasmas starting from the 
DSF, whereas in this work the dielectric function is the central quantity.

Having the thermal emission spectrum (\ref{eqn:def_gaunt}) and the
dynamic structure factor (\ref{eqn:skw_evan}) at our disposal, we are now
in the position to compare the signal from Thomson scattering to
the bremsstrahlung background thereby treating both processes on a common basis.

\section{Results}
\begin{figure}[t]
	\begin{center}
		\includegraphics[width=.5\textwidth,clip,angle=0]{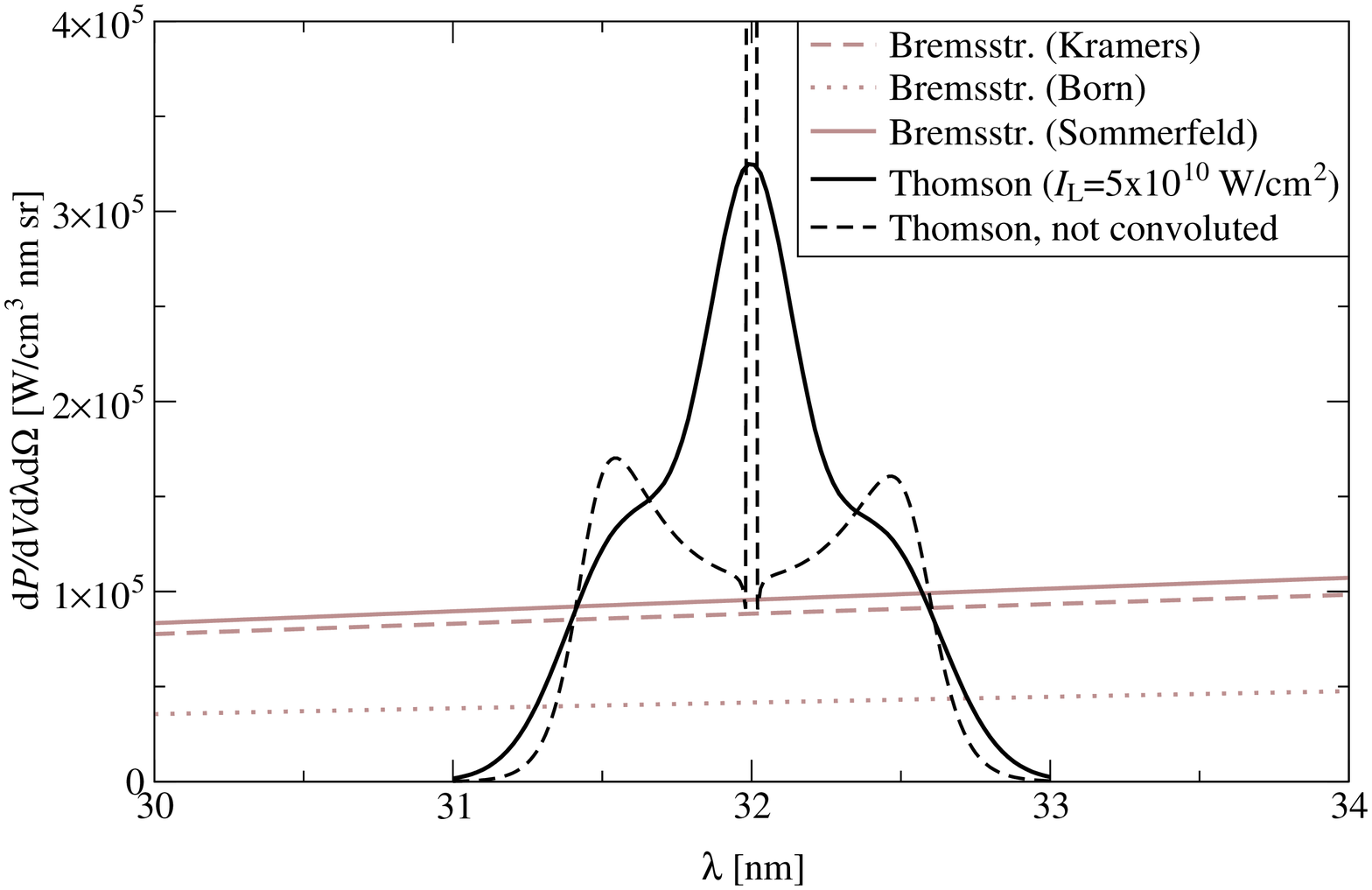}
		\caption{Emission power spectra for Thomson scattering and bremsstrahlung. The black dashed curve represents the unconvoluted Thomson spectrum. Parameters:
		$k_\mathrm{B}T_\mathrm{e}=10\,\mathrm{eV}$, $n_\mathrm{e}=10^{20}\,\mathrm{cm}^{-3}$, laser wavelength $\lambda=32\,\mathrm{nm}$, scattering
		parameter $\alpha=1.25$, $Z_\mathrm{eff}=1$~.}
		\label{fig:spec_320H120A-2_t10_n20}
	\end{center}
\end{figure}

\begin{figure}[t]
	\begin{center}
		\includegraphics[width=.5\textwidth,clip,angle=0]{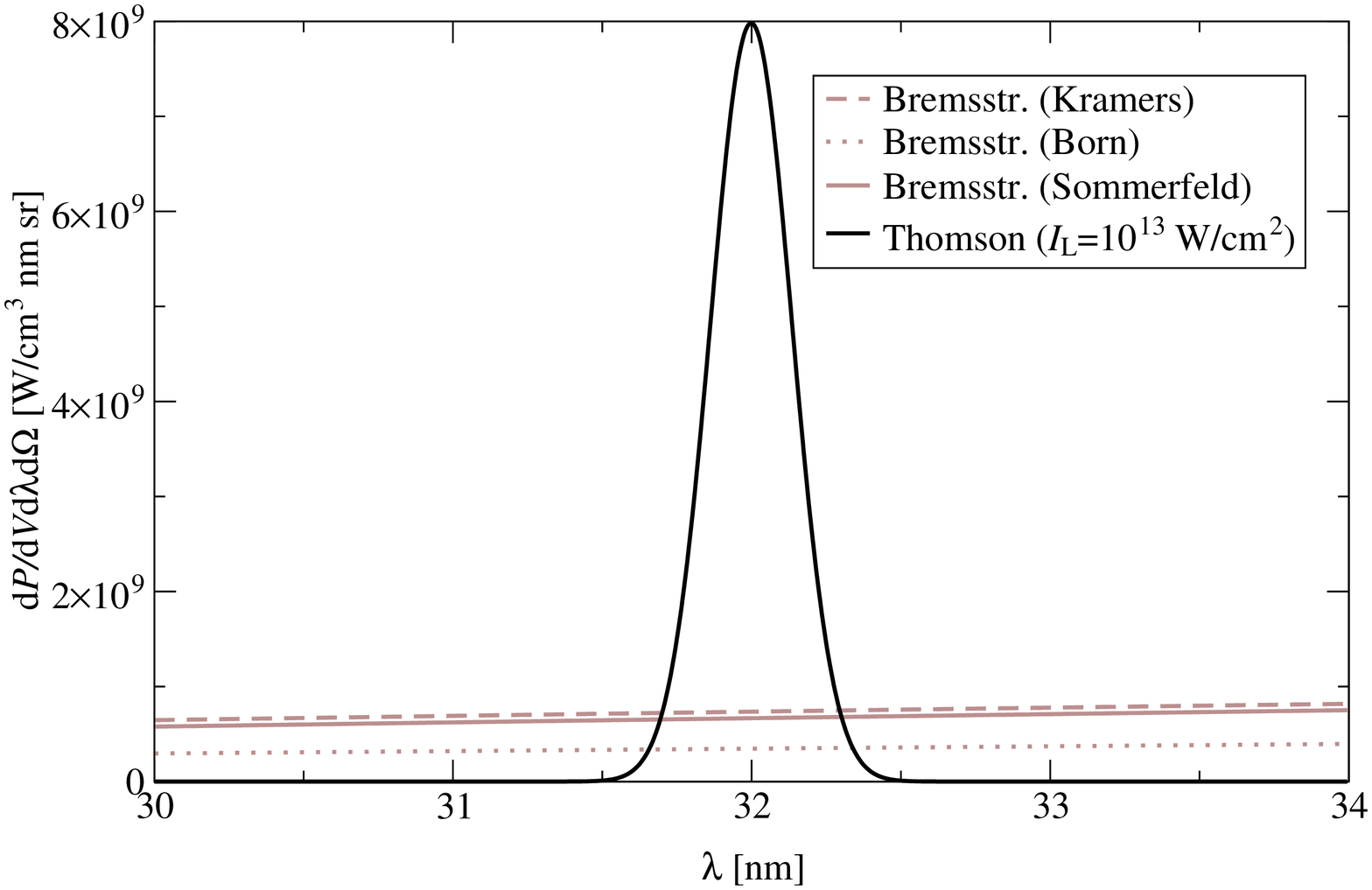}
		\caption{Emission power spectra for Thomson scattering and bremsstrahlung. Parameters:
		$k_\mathrm{B}T_\mathrm{e}=10\,\mathrm{eV}$, $n_\mathrm{e}=10^{22}\,\mathrm{cm}^{-3}$, laser wavelength $\lambda=32\,\mathrm{nm}$, scattering
		parameter $\alpha=12.5$, $Z_\mathrm{eff}=1$~.}
		\label{fig:spec_320H120A-2_t10_n22}
	\end{center}
\end{figure}

\begin{figure}[t]
	\begin{center}
		\includegraphics[width=.5\textwidth,clip,angle=0]{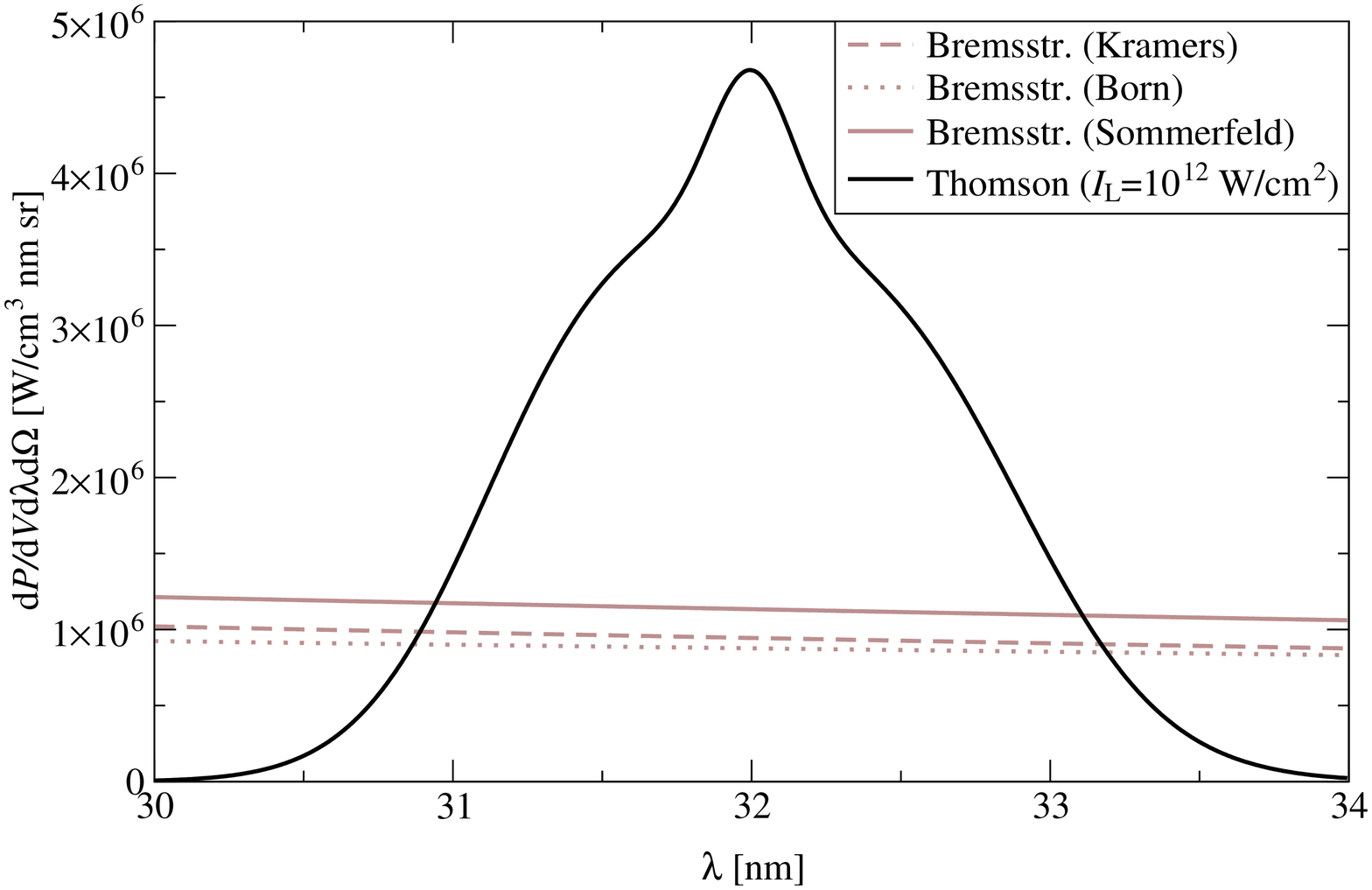}
		\caption{Emission power spectra for Thomson scattering and bremsstrahlung. Parameters:
		$k_\mathrm{B}T_\mathrm{e}=50\,\mathrm{eV}$, $n_\mathrm{e}=10^{20}\,\mathrm{cm}^{-3}$, laser wavelength $\lambda=32\,\mathrm{nm}$, scattering
		parameter $\alpha=0.56$, $Z_\mathrm{eff}=1$~.}
		\label{fig:spec_320H120A-2_t50_n20}
	\end{center}
\end{figure}

\begin{figure}[t]
	\begin{center}
		\includegraphics[width=.5\textwidth,clip,angle=0]{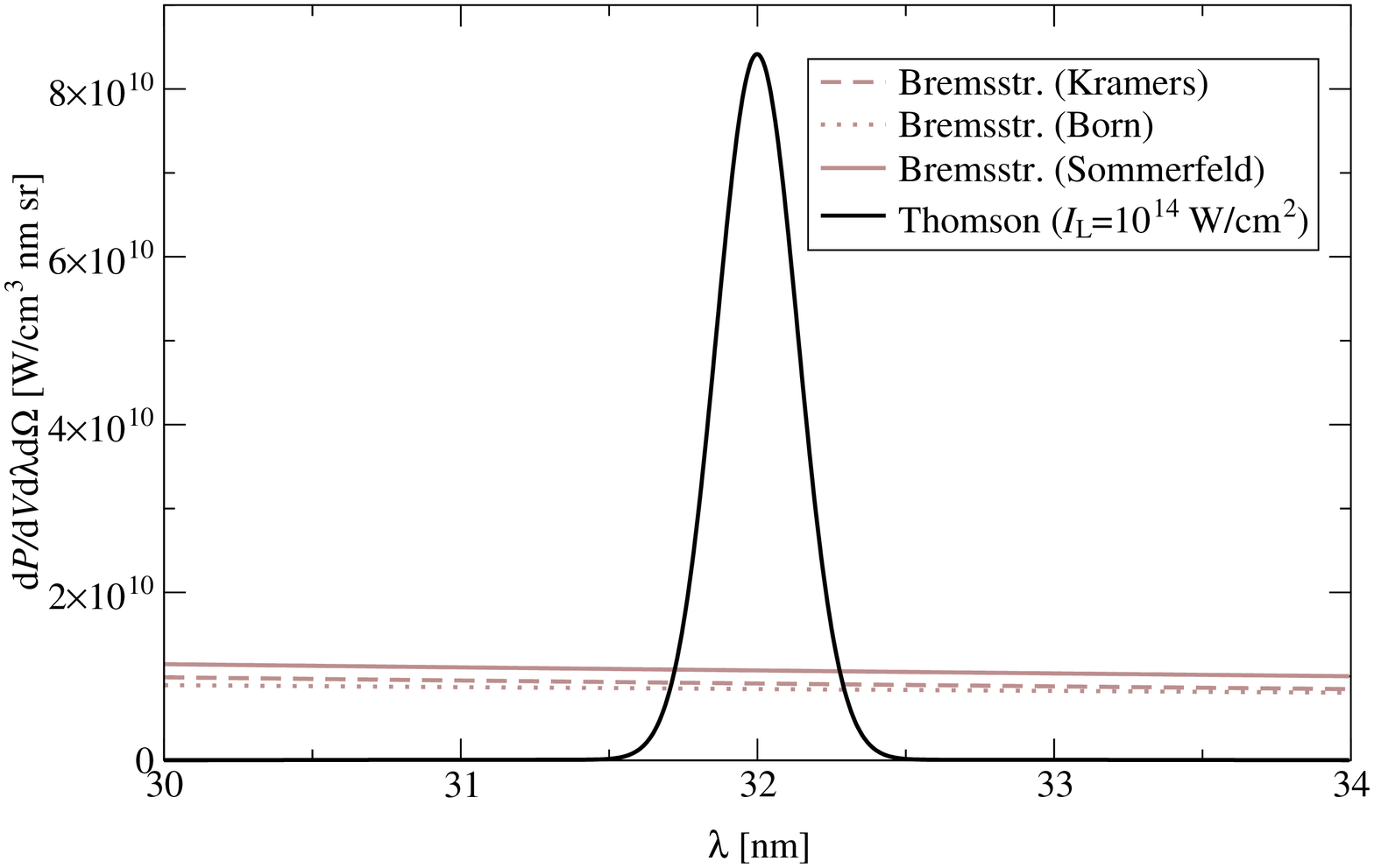}
		\caption{Emission power spectra for Thomson scattering and bremsstrahlung. Parameters:
		$k_\mathrm{B}T_\mathrm{e}=50\,\mathrm{eV}$, $n_\mathrm{e}=10^{22}\,\mathrm{cm}^{-3}$, laser wavelength $\lambda=32\,\mathrm{nm}$, scattering
		parameter $\alpha=5.6$, $Z_\mathrm{eff}=1$~.}
		\label{fig:spec_320H120A-2_t50_n22}
	\end{center}
\end{figure}

We now compare power spectra for Thomson scattering and
bremsstrahlung emission.
Figures \ref{fig:spec_320H120A-2_t10_n20} - \ref{fig:spec_320H120A-2_t50_n22} show results 
for different combinations of electron density and temperature for a hydrogenic plasma. We consider 
densities $n_\mathrm{e}=10^{20}\,\mathrm{cm}^{-3}$ and $10^{22}\,\mathrm{cm}^{-3}$ and temperatures of $10\,\mathrm{eV}$ and $50\,\mathrm{eV}$ as 
examples for possible experimental conditions with cryogenic targets. Furthermore, backscattering geometry (scattering angle $120^\circ$) is
chosen and a VUV laser wavelength of $\lambda=32\,\mathrm{nm}$ is assumed.

To model the detector, we assume a detector function
$G(\lambda )$ of
gaussian shape, its width given by the
relative spectral bandwidth $\Delta\lambda/\lambda=\Delta\omega/\omega$ 
\cite{bald:rev.sci.inst02}
\begin{equation}
	G(\lambda)=\frac{1}{\sqrt{2\pi\sigma^2}}\mathrm{e}^{-\lambda^2/2\sigma^2}~,\quad
	\sigma=0.425 \Delta\lambda~.
	\label{eqn:gaussian_sf}
\end{equation}
For the case of bremsstrahlung,
we neglect the effect of the finite bandwidth of the
detector, since the corresponding spectrum is only slowly 
varying with frequency.
The effective ion charge $Z_\mathrm{eff}$, calculated with COMPTRA04 \cite{kuhl:cpp05}
is close to $Z_\mathrm{eff}=1$ in the case of hydrogen for the present values of
electron density and temperature. 
 
In Fig.~\ref{fig:spec_320H120A-2_t10_n20}, the black dashed curve represents the 
pure Thomson signal, i.e. no convolution with the detector function $G(\lambda)$ has been performed. 
For the present parameters $n_\mathrm{e}=10^{20}\,\mathrm{cm}^{-3}$, and $k_\mathrm{B}T_\mathrm{e}=10\,\mathrm{eV}$, the Thomson spectrum contains a very
narrow ion peak, situated at the laser wavelength, and two satellites, which can be identified as electronic plasmon resonances. 
The central peak dominates these electronic resonances by a factor of 10, 
approximately.
The solid black curve is obtained by convolution of the 
pure Thomson signal (dotted black curve) with the
detector function.
The detector resolution is set to $\Delta\lambda=3.2\,\mathrm{nm}$, which corresponds to $1\%$ of the central wavelength.
Due to the broad detector function, the central peak is lowered and broadened such 
that the plasmon resonances 
do not show up as separate structures anymore. Instead, they only provide
sidewings in the spectrum.
In Fig.~\ref{fig:spec_320H120A-2_t10_n22} ($n_\mathrm{e}=10^{22}\,\mathrm{cm}^{-3}$, and $k_\mathrm{B}T_\mathrm{e}=10\,\mathrm{eV}$), the electronic peaks are
totally absorbed in the central peak. In this case, the ion peak in the unconvoluted
signal (not shown) is several orders of magnitude larger than the electronic resonances, which therefore do not contribute to the convolution integral (\ref{eqn:sig_thomson}). 
Fig.~\ref{fig:spec_320H120A-2_t50_n20} shows spectra for $n_\mathrm{e}=10^{20}\,\mathrm{cm}^{-3}$, and $k_\mathrm{B}T_\mathrm{e}=50\,\mathrm{eV}$. At these parameters,
one gets $\alpha=0.56<1$ for the 
scattering parameter, we are in the noncollective regime.
Already the unconvoluted Thomson signal is free of sharp electronic resonances, only
two shoulders appear in the broad, noncollective signal. 
Finally, in 
Fig.~\ref{fig:spec_320H120A-2_t50_n22} ($n_\mathrm{e}=10^{22}\,\mathrm{cm}^{-3}$, and $k_\mathrm{B}T_\mathrm{e}=50\,\mathrm{eV}$), we have the same structure as
in Fig.~\ref{fig:spec_320H120A-2_t10_n22}, no plasmon peaks are visible due to
the dominance of the ion feature. 

In Figures \ref{fig:spec_320H120A-2_t10_n20} - \ref{fig:spec_320H120A-2_t50_n22}, 
the intensity of the laser is chosen such that the central peak is situated clearly 
above the bremsstrahlung level.
Three approximations for bremsstrahlung are shown, namely Kramers formula (\ref{eqn:emis_kramers}), Born approximation (\ref{eqn:emis_cl_K0}) and
Sommerfeld's formula (\ref{eqn:somm1}). Born approximation gives larger deviations from Kramers result than the Sommerfeld expression.
This corresponds to the behaviour of the Gaunt factor shown in 
Fig.~\ref{fig:gaunt}: The Sommerfeld result is always closer to unity than Born approximation.

In the following, we evaluate the threshold intensity $I_\mathrm{thresh}(\omega)$ defined in Eq.~(\ref{eqn:thresh_int}) using
Eqs.~(\ref{eqn:sig_thomson}), (\ref{eqn:thoms_skw}), and (\ref{eqn:skw_evan}) for the Thomson power spectrum 
and Eq.~(\ref{eqn:def_gaunt}) with the Gaunt factor in either Born approximation (\ref{eqn:gaunt_K0}) or
t-matrix approximation (\ref{eqn:somm1}) for the bremsstrahlung power
spectrum. 

\begin{figure}[ht]
	\begin{center}
		\includegraphics[width=0.5\textwidth,clip,angle=0]{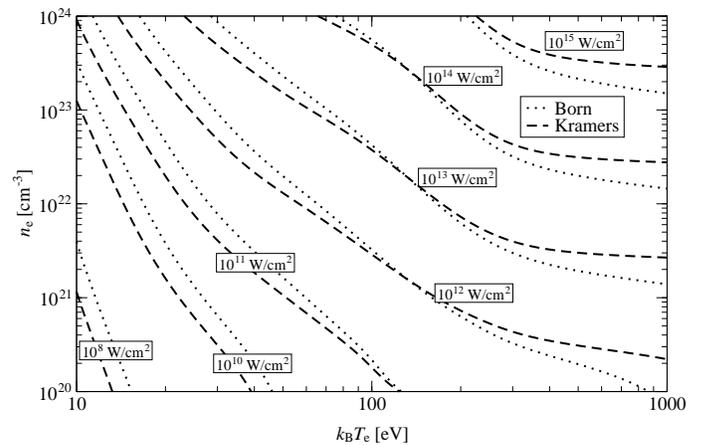}
		\caption{Threshold curves in the density-temperature plane for Al. Parameters:
		laser wavelength $14.7\,\mathrm{nm}$, $\theta=20^\circ$, $\Delta\lambda/\lambda=10^{-4}$, $T_\mathrm{e}/T_\mathrm{i}=2$.
		}
		\label{fig:thresh_147Al20A-4}
	\end{center}
\end{figure}

\begin{figure}[ht]
	\begin{center}
		\includegraphics[width=0.5\textwidth,clip,angle=0]{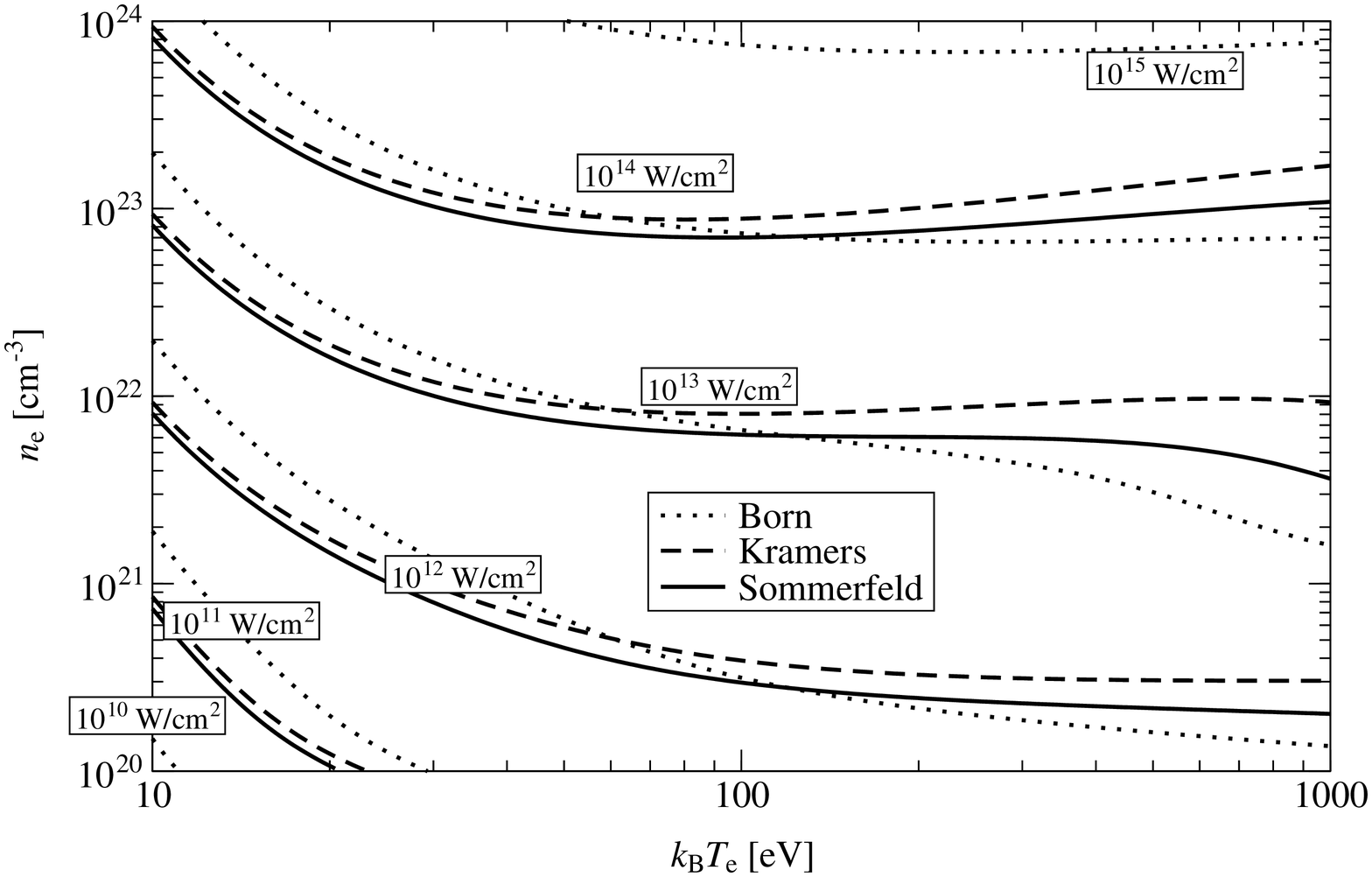}
		\caption{Threshold curves in the density-temperature plane for H. Parameters:
		laser wavelength $32\,\mathrm{nm}$, $\theta=120^\circ$, $\Delta\lambda/\lambda=10^{-2}$, $T_\mathrm{e}/T_\mathrm{i}=2$.}
		\label{fig:thresh_320H120A-2}
	\end{center}
\end{figure}

\begin{figure}[ht]
	\begin{center}
		\includegraphics[width=0.5\textwidth,clip,angle=0]{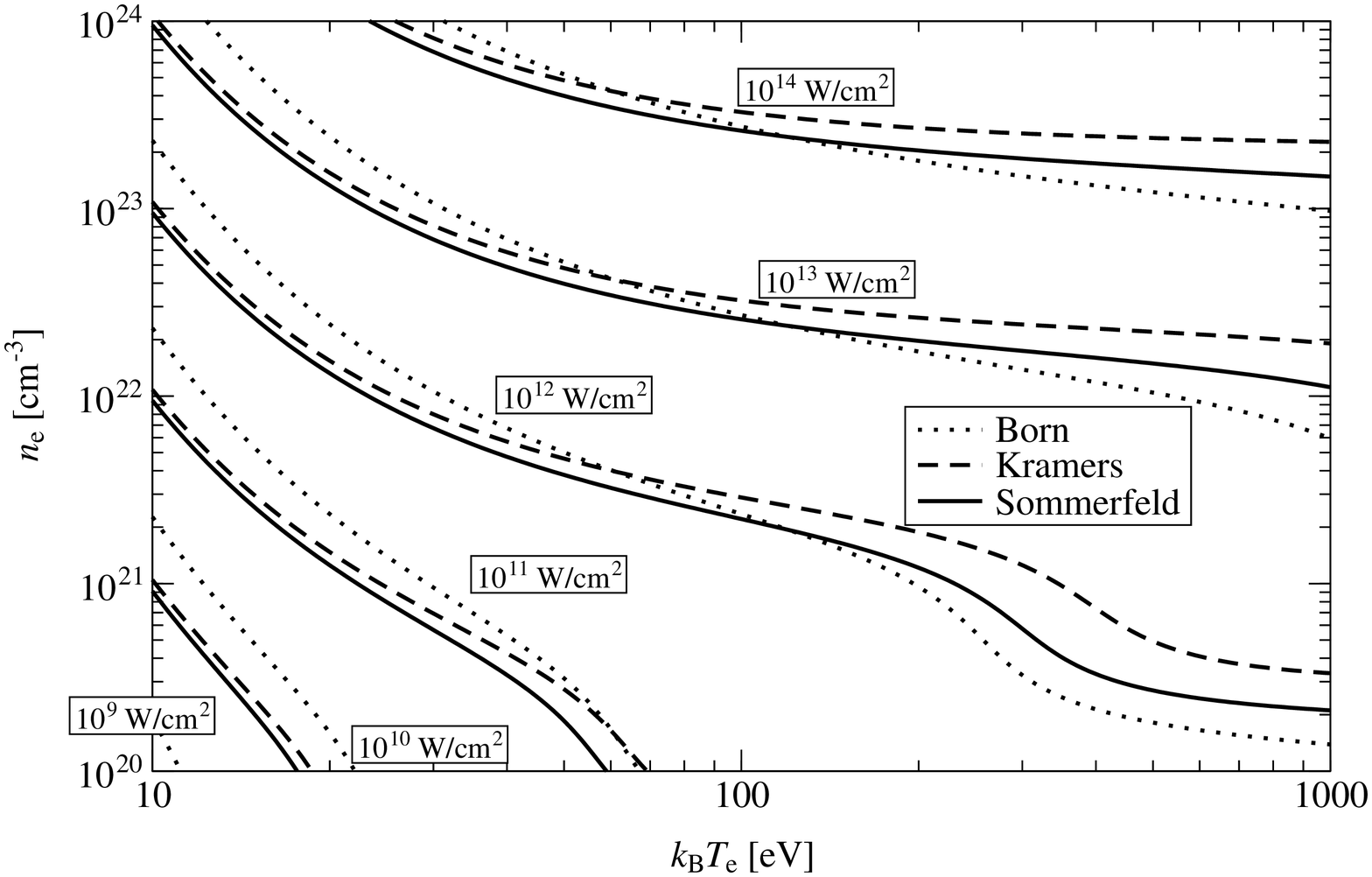}
		\caption{Threshold curves in the density-temperature plane for H. Parameters:
		laser wavelength $32\,\mathrm{nm}$, $\theta=120^\circ$, $\Delta\lambda/\lambda=10^{-4}$, $T_\mathrm{e}/T_\mathrm{i}=2$.}
		\label{fig:thresh_320H120A-4}
	\end{center}
\end{figure}

\begin{figure}[ht]
	\begin{center}
		\includegraphics[width=0.5\textwidth,clip,angle=0]{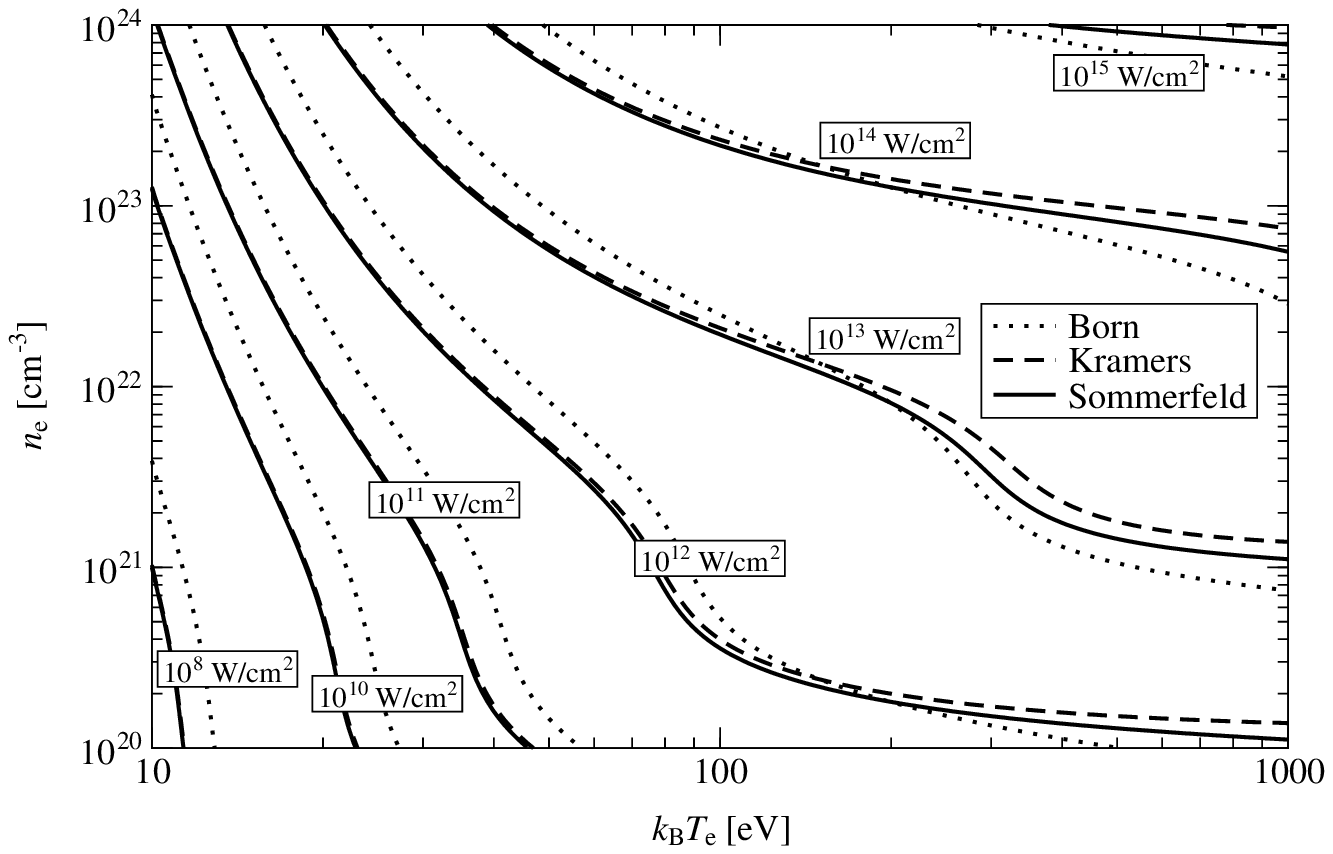}
		\caption{Threshold curves in the density-temperature plane for H. Parameters:
		laser wavelength $13.0\,\mathrm{nm}$, $\theta=120^\circ$, $\Delta\lambda/\lambda=10^{-4}$, $T_\mathrm{e}/T_\mathrm{i}=2$.}
		\label{fig:thresh_130H120A-4}
	\end{center}
\end{figure}

Figures \ref{fig:thresh_147Al20A-4} - \ref{fig:thresh_130H120A-4} show contour plots of the threshold intensity in part of the  
$n_\mathrm{e}-T_\mathrm{e}$-plane. 
In Fig.~\ref{fig:thresh_147Al20A-4} we compare Kramers result (dashed curves) to
Born approximation (dotted curves) assuming aluminum as target material, scattering angle $\theta=20^\circ$, $\Delta\lambda/\lambda=10^{-4}$. In this case,
the effective ion charge $Z_\mathrm{eff}$ is calculated using Thomas-Fermi theory \cite{more:ir_llnl} as was also done by Baldis \textit{et al.} in Ref.~\cite{bald:rev.sci.inst02}, who used the same set of parameters. 
Their results are
reproduced by using Kramers formula for bremsstrahlung (dashed curve). 

For low temperatures, where the Gaunt factor in 
Born approximation is smaller than unity at the considered photon energy, cf. Fig.~\ref{fig:gaunt}, higher densities
are accessible as compared to Kramers result. For high temperatures, 
the opposite becomes true, $\bar g^\mathrm{B}(\omega)>1$ leads to lower accessible 
densities.\\ 
In figures \ref{fig:thresh_320H120A-2} - \ref{fig:thresh_130H120A-4}, three approximations have been calculated for the bremsstrahlung level. Besides Kramers formula (dashed curves) we show results for
Born approximation (dotted curves) and Sommerfeld's formula (solid curves).
Comparing the Sommerfeld result to Born approximation, it can be noted that Born approximation gives larger corrections, 
while Sommerfeld's theory leads to smaller deviations from Kramers result.
This is one important result of this work: Taking into account quantum effects in
a rigorous way via Sommerfeld's expression for the Gaunt factor leads only to small 
corrections of threshold intensities, while 
Born approximation tends to overestimate these effects. This underlines the 
importance to go beyond Born approximation.
For moderate and high temperatures, the Sommerfeld result lies systematically below the Kramers result, due to the increasing Gaunt factor at high 
temperatures.

Furthermore, we investigated the influence of other experimental parameters, namely the laser wavelength, the material, and the spectral bandwidth of the 
spectrometer. The latter parameter turns out to be of great importance: By comparison of Fig.~\ref{fig:thresh_320H120A-2} with Fig.~\ref{fig:thresh_320H120A-4},
one observes that notably higher densities are accessible in the case of small spectral bandwidth ($\Delta\lambda/\lambda=10^{-4}$ in Fig.~\ref{fig:thresh_320H120A-4})
than for the relatively large bandwidth ($\Delta\lambda/\lambda=10^{-2}$ in Fig.~\ref{fig:thresh_320H120A-2}).

Comparing  Fig.~\ref{fig:thresh_147Al20A-4} to Fig.~\ref{fig:thresh_320H120A-4}, the influence of the $Z$-number of the material becomes apparent:
Low $Z$-materials produce less bremsstrahlung than high $Z$-elements due to the $Z^2$-proportionality of the bremsstrahlung cross section, 
cf. Eq.~(\ref{eqn:emis_kramers}). 
On the other hand, the $Z$-dependance of the dynamical structure factor largely cancels out for the free electron part of the 
structure factor while it is roughly $Z/(1+Z)^2$ for the ionic part, cf. Eq.~(\ref{eqn:skw_evan}). 

Finally, important differences are noted upon changing the laser wavelength from
$32\,\mathrm{nm}$ (Fig.~\ref{fig:thresh_320H120A-4}) to $13\,\mathrm{nm}$ (Fig.~\ref{fig:thresh_130H120A-4}) especially at low temperatures.
Both data sets have been
calculated using the same spectral resolution ($\Delta\lambda/\lambda=10^{-4}$) and material (H). Looking at Fig.~\ref{fig:thresh_130H120A-4}
($\lambda=13\,\mathrm{nm}$) at low temperatures
the threshold contours are almost parallel to the density axis.  
For higher temperatures, both wavelengths $32\,\mathrm{nm}$ and $13\,\mathrm{nm}$ give nearly equal threshold intensities, which do not depend on temperature.
This can be understood from the bremsstrahlung spectrum, which
becomes nearly independent of frequency and temperature at low $\hbar\omega/k_\mathrm{B}T$.



\begin{table}[ht]
\caption{Experimental parameters and resulting threshold
intensities for different electron temperatures and laser wavelengths. Fixed parameters:
$Z=1$, $n_\mathrm{e}=10^{21}\,\mathrm{cm}^{-3}$. }
\small
	\begin{tabular}{|c|c|c|c|c|c|c|}
		\hline
		$k_\mathrm{B}T_\mathrm{e}[\mathrm{eV}]$ &
		$\lambda$[nm]&
		$\theta$&
		$\Delta\lambda/\lambda$&
		$\lambda_\mathrm{pl}$[nm]&
		$I_\mathrm{thresh}[\mathrm{W/cm}^2]$&$\bar g^\mathrm{S}_\mathrm{ff}$
		\\
		\hline
		$10$&$32$&$120$&$10^{-2}$&$33.1$&$1.37\cdot 10^{12}$&1.14\\
		$20$&$32$&$120$&$10^{-2}$&$33.2$&$4.92\cdot 10^{12}$&1.17\\
		$50$&$32$&$120$&$10^{-2}$&$33.4$&$9.29\cdot 10^{12}$&1.22\\
		\hline
		$10$&$13$&$120$&$10^{-4}$&$13.3$&$2.95\cdot 10^{9}$&1.03\\
		$20$&$13$&$45$&$10^{-4}$&$13.2$&$1.55\cdot 10^{11}$&1.04\\
		$50$&$13$&$45$&$10^{-4}$&$13.2$&$3.49\cdot 10^{12}$&1.06\\
		\hline
\end{tabular}
\label{tab:parameters}
\end{table}

Since the information about temperature and density is stored in the position and height of the plasmon resonances, we will now focus on
experimental conditions to be met in order to separate the plasmon peak from bremsstrahlung background. Results for the threshold intensity are given in Tab.~\ref{tab:parameters}.
The electron density is set to $n_\mathrm{e}=10^{21}\,\mathrm{cm}^{-3}$, for $n_\mathrm{e}=10^{20}\,\mathrm{cm}^{-3}$ and $10^{22}\,\mathrm{cm}^{-3}$ see
figures \ref{fig:spec_320H120A-2_t10_n20} - \ref{fig:spec_320H120A-2_t50_n22}.
Two wavelengths are considered, $\lambda=32\,\mathrm{nm}$, the 
momentary VUV-FEL wavelength at DESY, and $\lambda=13\,\mathrm{nm}$, envisaged wavelength in the near future.
Since the latter wavelength allows for very efficient x-ray optics to be applied, the spectral resolution has been reduced to $\Delta\lambda/\lambda=10^{-4}$, while
$32\,\mathrm{nm}$ wavelength allows only for $\Delta\lambda/\lambda=10^{-2}$. Finally the scattering angle has been chosen such that the plasmon peak is on the one hand 
well pronounced but on the other hand shifted far enough from the central peak as to be resolved by the spectrometer. For detailed discussion of
the Thomson scattering spectrum at various plasma parameters, we refer to Refs.~\cite{hoel:epjd04, redm:ieee05}. The rightmost column of Tab.~\ref{tab:parameters} gives the Gaunt factor in Sommerfeld approximation (Eq.~(\ref{eqn:somm1})). Note that in the case of the $32\,\mathrm{nm}$ wavelength, 
the threshold intensity is increased by $14-22\%$ due to the Gaunt factor.

\section{Conclusion}
In this work we have studied Thomson scattering and bremsstrahlung emission 
in warm dense matter in the context of plasma diagnostic experiments
to be performed in near
future. Out of these two competing, Thomson scattering
is supposed to serve as a probe for plasma parameters, but
also bremsstrahlung gives an important contribution to the
photon yield from highly ionized plasmas. 
Thus, experimental conditions
have to be determined that allow for a maximum 
signal-to-background ratio. Here, we focused on the laser intensity. To this purpose, expressions for the
Thomson signal, which is given by the dynamical structure
factor of the plasma as well as for the bremsstrahlung 
spectrum have been derived from a common starting point,
namely linear response theory. 
This approach allows for a systematic treatment of medium and quantum
effects, such as dynamical screening and strong collisions.

We have applied our formulas to determine threshold intensities
for the external photon source (FEL) as function of
density and temperature as well as their dependence
on further parameters, namely the laser wavelength, the
spectral resolution of detectors, and the material. 
In the discussion we focused on the bremsstrahlung spectrum.
It was shown that Born approximation overestimates the effect of collisions, i.e. it leads to
relatively high Gaunt factors if compared to Sommerfeld's expression. Sommerfeld's formula (\ref{eqn:somm1}) gives the correct
Gaunt factor in a weakly coupled plasma. 
At temperatures $k_\mathrm{B}T_\mathrm{e}\geq20\,\mathrm{eV}$ threshold intensities calculated with the Gaunt factor in Sommerfeld approximation 
are larger than those obtained by
Baldis \textit{et al.} in Ref.~\cite{bald:rev.sci.inst02}, who used Kramers expression, i.e. $\bar g_\mathrm{ff}=1$.

The Thomson signal was analysed within RPA which gives the contribution in lowest order of density. For densities investigated here, collisions are also
relevant for the Thomson signal and lead to a pronounced change of the electronic part in $S(\boldsymbol{k},\lambda)$. This was investigated in Ref.~\cite{thie:j.physA06}.
However, the account for collisions does not alter the results for the threshold intensities performed in this work, since the Thomson scattering signal was evaluated at the
laser wavelength, were the ionic feature of the DSF dominates.

Finally, we have demonstrated, that Thomson scattering can easily overcome
the bremsstrahlung background, if laser intensities of $10^{8}\,\mathrm{W}/\,\mathrm{cm}^2-10^{13}\,\mathrm{W}/\,\mathrm{cm}^2$ are provided. From this point of view,
Thomson scattering experiments 
for plasma diagnostics using VUV-FEL radiation 
seem feasible.

\begin{acknowledgments}
This work has been supported by the Virtual Institute
VH-VI-104 \textit{Plasma Physics Research using FEL 
Radiation} of the Helmholtz Society and the 
DFG-Sonderforschungsbereich 652 \textit{Starke Korrelationen und kollektive
Ph\"anomene im Strahlungsfeld:
Coulombsysteme, Cluster und Partikel}. We gratefully acknowledge A. H\"{o}ll and 
J. Chihara for
stimulating discussion.
C.F. would like to thank DESY for hospitality and support during a two week visit.
\end{acknowledgments}
\begin{appendix}
\section{\label{app:lrt}Correlation functions and Green functions}
Correlation functions for two observables $A,B$ 
are defined according to 
\begin{eqnarray}
       (A;B) &=& {1 \over \beta} \int\limits_0^\beta \mathrm{d} \tau\, 
        {\rm Tr}[A(-i  \hbar \tau) B^\dagger \rho_0]
        \;\;\;, \nonumber\\
        \langle A;B \rangle_{\omega +i \eta}  &=& 
        \int\limits_0^\infty \mathrm{d} 
        t\,\mathrm{e}^{i (\omega +i \eta)t}\, 
        (A(t);B)~,
\end{eqnarray}
$\rho_0$ is the equilibrium statistical operator.
The time dependence of observables is taken in the Heisenberg picture, using the system Hamiltionian
\begin{equation}
	H\!=\!\sum_{\boldsymbol{p},c}\!\frac{\hbar^2p^2}{2m_c}a^\dagger_{\boldsymbol{p},c}a^{}_{\boldsymbol{p},c}+\frac{1}{2}\!\!\sum_{\boldsymbol{p}\boldsymbol{k}\boldsymbol{q},cd}\!\!\!V_{cd}(\boldsymbol{q})a^\dagger_{\boldsymbol{p}+\boldsymbol{q},c}a^\dagger_{\boldsymbol{k}-\boldsymbol{q},d}a^{}_{\boldsymbol{k},d}a^{}_{\boldsymbol{p},c}~.
	\label{eqn:hamitonian}
\end{equation}
Performing integration by parts, the current-current correlation function (\ref{eqn:alpha_realjj}) can be expressed through
a force-force correlation function as
\begin{widetext}
\begin{equation}
        \chi^{-1}(\boldsymbol{k}, \omega) =  {i 
        \over \beta \Omega_0} {\omega q^2 \over
        k^2} {1 \over ( J^z_k ; J^z_k)^2 } \left[-i 
        \omega ( J^z_k ; J^z_k) +
        \langle \dot J^z_k ; \dot J^z_k \rangle_{\omega + i \eta} - {\langle
        \dot J^z_k ; J^z_k \rangle_{\omega + i  \eta} 
        \langle J^z_k ; \dot
        J^z_k \rangle_{\omega + i  \eta} 
        \over \langle J^z_k ; J^z_k
        \rangle_{\omega + i  \eta}} \right]~,
\end{equation}
\end{widetext}
with  
$ \dot J^z_{0,{\rm e}} = i\big[H,J^z_{0,{\rm
            e}}\big]/\hbar$.

Also, it is convenient to introduce a 
generalized collision frequency $\nu(\omega)$
in analogy to the Drude relation \cite{rein:pre00}
$ \epsilon(\omega)  =  1- \omega_{\rm pl}^2/(\omega^2+i\omega\nu(\omega))$
where $\omega_{\rm pl}^2=\sum_c\, n_ce_c^2/(\epsilon_0m_c)$ 
is the squared plasma frequency. \\
By comparison with Eq.~(\ref{eqn:eps_chi}) using $( J^z_k ; J^z_k)=\epsilon_0\omega^2_\mathrm{pl}/ \beta\Omega_0$, we establish an expression 
for the collision frequency in terms of correlation functions
\begin{multline}
        \label{Def:nu}
        \nu(\omega) = \frac{\beta \Omega_0}{\epsilon_0\omega_{\rm pl}^
        {2}}\,
        \lim_{k \to 0}\,\bigg[
        \big\langle \,\,\dot{\!\!J}_{k}^{z} , \,\,\dot{\!\!J}_{k}^{z}\big\rangle_{\omega+i \eta}\\
	-\frac{\langle \dot J^{z}_{k} ; J^{z}_{k} \rangle_{\omega + i \eta}\, 
	\langle J^{z}_{k} ; \dot J^{z}_{k} \rangle_{\omega + i  \eta}}
	{ \langle J^{z}_{k} ; J^{z}_{k} \rangle_{\omega + i  \eta}} 
        \bigg]~.
\end{multline}
Further details can be found in Ref.~\cite{rein:pre00}.
Making use of Eq.~(\ref{eqn:alpha_eps}), 
the absorption coefficient can be expressed as
\begin{equation}
        \alpha(\omega)= \frac{\omega_{\rm pl}^{2}}{c}
        \frac{{\rm Re}\;\nu(\omega)}{(\omega_{}^{2}-2 \omega\, 
        {\rm Im}\,\nu(\omega) + |\nu(\omega)|^2)
        n(\omega)}\;\;\;.
\end{equation}
In the high frequency limit $\omega \gg \omega_{\rm pl}$,
the index of refraction is unity, the imaginary part of the collision frequency tends to zero and the collision frequency is small
compared to the frequency $\omega$. Then, we can consider the
approximation $\alpha(\omega)  =  \omega_{\rm pl}^2\mbox{Re}\, \nu(\omega)/\omega^2c 
        =\beta \Omega_0 {\rm Re}\, \big\langle \,\,\dot{\!\!J}_{0}^{z} , \,\,\dot{\!\!J}_{0}^{z}
        \big\rangle_{\omega+i \eta} /c \epsilon_0 \omega^2$,
where the collision frequency is given in the form of a force-force
correlation function, cf.~Ref.~\cite{rein:pre00}. Thus, the absorption
coefficient is directly proportional to the 
real part of the force-force correlation function, which itself
can be determined using perturbation theory. 

The relation to the thermodynamic Green function of two observables is given
by
\begin{equation}
	\langle A;B \rangle_{\omega +i \eta}=\frac{i}{\pi\beta}\int_{-\infty}^\infty\frac{\mathrm{d}
	\bar\omega}{\bar\omega}\frac{1}{\omega+i\eta-\bar\omega}\mathrm{Im}\,G_{AB}(\bar\omega+i\eta)~.	
	\label{eqn:def_greenf}
\end{equation}
Using Dirac's identity
one obtains
$\mathrm{Re}\,\langle A;B \rangle_{\omega +i \eta}=\mathrm{Im}\,G_{AB}(\omega+i\eta)/\omega\beta$~,
which directly leads to Eq.~(\ref{eqn:alpha_realjj}).

\section{\label{app:chi_rpa}The RPA-susceptibility}
Eq.~(\ref{eqn:chi_rpa_vsc_text}) is obtained from the exact relation for 
the response function $\chi_{cc'}(\boldsymbol{k},\omega)$, 
\begin{equation}
	\chi_{cc'}(\boldsymbol{k},\omega)=\chi_{c}^0(\boldsymbol{k},\omega)\delta_{cc'}+\sum_d \chi_c^0(\boldsymbol{k},\omega)V_{cd}^\mathrm{sc}(\boldsymbol{k},\omega)\chi_{dc'}(\boldsymbol{k},\omega)~,
	\label{eqn:dyson_chi}
\end{equation}
by truncation after the first iteration, i.e. $\chi_{dc'}=\chi^0_{dc'}$ in the second term of Eq.~(\ref{eqn:dyson_chi}).
Insertion of the screened potential Eq.~(\ref{eqn:dyson_Vsc_text}) yields the closed equation
\begin{widetext}
\begin{align}
\begin{split}
		\chi^\mathrm{RPA}_{cc'}(\boldsymbol{k},\omega)&=\chi_c^0(\boldsymbol{k},\omega)\delta_{cc'}+\chi_{c}^0(\boldsymbol{k},\omega)\left[%
		V_{cc'}(\boldsymbol{k})+\sum_d V_{cd}(\boldsymbol{k})\chi_d^0(\boldsymbol{k},\omega)V^\mathrm{sc}_{dc'}(\boldsymbol{k},\omega)\right]\chi_{c'}^0(\boldsymbol{k},\omega)\\
	&=\chi_c^0(\boldsymbol{k},\omega)\delta_{cc'}+\sum_d \chi_c^0(\boldsymbol{k},\omega)V_{cd}(\boldsymbol{k})\chi_{dc'}^\mathrm{RPA}(\boldsymbol{k},\omega)~.
\label{eqn:dyson_chi_rpa_2}
\end{split}
\end{align}
\end{widetext}

For a two component plasma, i.e. an electron-ion plasma
$c=\mathrm{e},\mathrm{i}$, 
we obtain by matrix inversion

\begin{eqnarray}
	\nonumber
	\chi^\mathrm{RPA}_\mathrm{ee}&=&\frac{\chi_\mathrm{e}^0(1-\chi_\mathrm{i}^0V_\mathrm{ii})}{1-\chi_\mathrm{e}^0V_\mathrm{ee}-\chi_\mathrm{i}^0V_\mathrm{ii}+\chi_\mathrm{i}^0\chi_\mathrm{e}^0(V_\mathrm{ii}V_\mathrm{ee}-V_\mathrm{ei}V_\mathrm{ie})}~,\\
	\label{eqn:chi_rpa_ee}\\
	\nonumber
	\chi^\mathrm{RPA}_\mathrm{ie}
	&=&\frac{\chi_\mathrm{e}^0V_\mathrm{ei}\chi_\mathrm{i}^0}{1-\chi_\mathrm{e}^0V_\mathrm{ee}-\chi_\mathrm{i}^0V_\mathrm{ii}+\chi_\mathrm{i}^0\chi_\mathrm{e}^0(V_\mathrm{ii}V_\mathrm{ee}-V_\mathrm{ei}V_\mathrm{ie})}~.\\
\label{eqn:chi_rpa_ie}
\end{eqnarray}
$\chi_\mathrm{ei}^\mathrm{RPA}$ and $\chi_\mathrm{ii}^\mathrm{RPA}$ are obtained by interchanging indices i and e in Eq.~(\ref{eqn:chi_rpa_ie}) and Eq.~(\ref{eqn:chi_rpa_ee}), respectively.

The imaginary part of the electronic susceptibility in RPA, $\chi^\mathrm{RPA}_\mathrm{ee}(\boldsymbol{k},\omega)$ (Eq.~(\ref{eqn:chi_rpa_ee})) for a hydrogen plasma ($V_\mathrm{ee}=V_\mathrm{ii}=-V_\mathrm{ei}=-V_\mathrm{ie}\equiv V$) now evaluates to
\begin{multline}
	\mathrm{Im}\,\chi^\mathrm{RPA}_\mathrm{ee}(\boldsymbol{k},\omega)=\\
	\left|\frac{1-V(\boldsymbol{k})\chi_\mathrm{i}^0(\boldsymbol{k},\omega)}{1-V(\boldsymbol{k})\left[\chi_{\mathrm e}^0(\boldsymbol{k},\omega)+\chi_\mathrm{i}^0(\boldsymbol{k},\omega)\right]}\right|^2\mathrm{Im}\,\chi_\mathrm{e}^0(\boldsymbol{k},\omega)+\\
	\ \ +\left|\frac{V(\boldsymbol{k})\chi_\mathrm{e}^0(\boldsymbol{k},\omega)}{1-V(\boldsymbol{k})\left[\chi_\mathrm{e}^0(\boldsymbol{k},\omega)+\chi_\mathrm{i}^0(\boldsymbol{k},\omega)\right]}\right|^2\mathrm{Im}\,\chi_\mathrm{i}^0(\boldsymbol{k},\omega)~.
	\label{eqn:im_chi_rpa}
\end{multline}
This expression can also be derived from kinetic theory, i.e. the perturbative expansion of Vlasov's equation for a two component
plasma \cite{evan:rep.prog.phys69}. It was used in \cite{bald:rev.sci.inst02} to compare the Thomson signal to the
emission background caused by thermal bremsstrahlung photons.
Generalizing for different electron and ion temperatures $T_\mathrm{e}$ and $T_\mathrm{i}$ and inserting the explicit expression for
the response function in RPA (\ref{eqn:chi_0_W}), one obtains Eq.~(\ref{eqn:skw_evan}).
\end{appendix}

\end{document}